# Rock-bridge failure criteria

Fengchang Bu[1,2,3*], Michel Jaboyedoff[3], Lei Xue[4]

1. Department of Earth Science and Engineering, Imperial College London, London, UK
2. Department of Civil and Environmental Engineering, Massachusetts Institute of Technology, Cambridge, MA, USA
3. Institute of Earth Sciences, University of Lausanne, Lausanne, Switzerland
4. State Key Laboratory of Lithospheric and Environmental Coevolution, Institute of Geology and Geophysics, Chinese Academy of Sciences, Beijing, China

*Corresponding author (Fengchang Bu: fengchang.bu@unil.ch)

**Abstract**

The intact intervals between discontinuities, known as rock bridges, have long been interpreted as crucial in resisting rock mass instabilities like faulting, landslides, and earthquakes. Despite over half a century of research on rock bridges, little, if anything, is known about their explicit resistance roles, such as how much resistance a rock bridge can provide, how it fails, what the resulting failure morphology is, and how to incorporate it into the field. Here we propose rock-bridge failure criteria, which can derive the resistance magnitude and failure modes of rock bridges. Specifically, the failure criterion for rock bridges defined by non-overlapping discontinuities is based on identifying the intersection of stress paths on ligaments and Mohr's circle or failure envelope, while that with overlapping discontinuities uses cantilever beam simplification. The rock-bridge failure criteria are generally applicable across various discontinuity arrangements, boundary conditions, multiple rock bridges, infilled discontinuities, and field conditions. For validation, we compile an extensive laboratory database comprising published and original datasets. Comparisons between the experimental and theoretical rock-bridge resistance and failure modes demonstrate the feasibility and robustness of the rock-bridge failure criteria. Therefore, the rock-bridge failure criteria provide a fundamental solution for analysing rock bridges, advancing our understanding and management of the associated geological phenomena and rock engineering.

## 1. Introduction

Natural rock masses are always dissected by persistent or impersistent discontinuities.[1] Instabilities of the rock masses are much more likely to occur if the discontinuities are fully persistent, given the usually much higher stiffness and strength of the rock matrix than those of the discontinuities.[2] Therefore, a thorough understanding of the failure of intact intervals between discontinuities, known as rock bridges, is fundamental for interpreting rock mass instabilities.

Rock bridges are purely geological structures but have long been interpreted through a mechanical lens. Müller[3] and Terzaghi[4] pioneered the conceptualisation of rock bridges and suggested that they could be possible resistance loci for rock slope instabilities. Since then, rock bridges have been widely analysed in landslide cases, such as the 1991 Randa rockslide,[5] 2007 Chamousset collapse,[6] Rundefjellet rockslide,[7] 2017 Pizzo Cengalo rock avalanche,[8] and 2023 La Praz collapse.[9] Exposed to natural events or human activities, intact rock bridges can accumulate damage to fail, interconnecting impersistent discontinuities and forming slip surfaces to trigger landslides. Rock bridges have also attracted attention in other fields, including earthquakes,[10-12] tectonics,[13-15] underground mining,[16, 17] and tunnel excavations.[18, 19] Although rock bridges may be termed differently, for example, gaps, steps, stepovers, or cohesion-healed patches, they are generally interpreted as zones of considerable stress concentrations and critical factors to resist instabilities.

To elucidate the resistance role of rock bridges, several rock-bridge failure criteria have been proposed. Lajtai[20] subjected plaster rock bridges to direct shear under different confining pressures, found tension, shear, and compression failure mechanisms of rock bridges, and proposed corresponding failure criteria. Jennings[21] proposed a weighted average of proportions and properties of rocks and discontinuities under the Mohr–Coulomb failure criterion. Both Lajtai's criterion and Jennings's criterion have been refined by integrating various models over decades. For example, Liu and Xia[22] modified the strength criterion for tensile failure in Lajtai's criterion; Wang et al.[23] coupled Lajtai's criterion with the strength criterion for discontinuities proposed by Barton and Choubey[24]; Rafiei Renani et al.[25] attempted to couple Jaeger's single plane of weakness theory with Jennings's criterion, where rock bridges are treated as equivalent discontinuities with enhanced cohesion and friction; and Hu et al.[26] modified Jennings's criterion by considering the dilatancy angle and localised weakening. Despite Lajtai's criterion and Jennings's criterion, alongside their developments, remaining the dominant paradigms in the rock-bridge domain, fundamental deficiencies persist. Both Lajtai's criterion and Jennings's criterion are based on the concept of discontinuity persistence $K$, which has been acknowledged to possess limitations,[27] such as its restricted applicability to coplanarity,[28] neglect of progressive failure,[29] and omission of scale effects.[30] In addition,

Lajtai's criterion, developed from direct shear tests, has not been generalised to other stress regimes. Subsequent refinements remain constrained by the concept of $K$ and direct shear conditions. Furthermore, the discrete consideration of the three failure modes precludes its application to the mixed-mode failure prevalent in nature and experiments, thereby limiting its practical adoption. In contrast, Jennings's criterion focuses exclusively on shear failure, failing to account for tensile and mixed-mode failure mechanisms. It also suffers from a heavy reliance on the concept of $K$ and overlooks the stress-orientation dependence of rock-bridge behaviour. Subsequent modifications, while incorporating some factors like progressive fracturing, have failed to overcome these fundamental limitations.

Given these limitations, how should we analyse rock bridges? Elmo et al.[31, 32] realised that rock bridges are complex zones of influence where failure is governed by fracture propagation and coalescence. In this view, rock bridges are considered as products of the failure process itself, and their real appearance is exposed only after failure,[29, 33] except in relatively simple geological conditions.[34] Accordingly, Elmo[35] proposed the Bologna interpretation of rock bridges: 'Prior to failure it is not a matter of describing rock bridges as physical deterministic entities. There are only potential rock bridges which exist everywhere at once. Only by looking at a rock mass as it fails, we can summon the position of a rock bridge into existence, and therefore measure the rock bridge potential of a rock mass.' In this interpretation, the rock bridge potential represents the contribution of rock bridges to the rock mass strength. Its determination requires numerical simulations to reproduce the rock mass failure process. Clearly, such interpretation and the metric refer to the rock bridge with mechanical significance; however, the relationship between rock-bridge configurations and mechanical behaviour remains elusive.

To establish this relationship, it is imperative to clarify the complexity of rock-bridge problems: (1) The diversity of rock-bridge configurations. Natural impersistent discontinuities can be coplanar, oblique, or overlapping, with various dimensions, resulting in various rock-bridge configurations and, consequently, various mechanical responses under specific boundary conditions. (2) Stress-orientation dependence. Even for a given rock-bridge configuration, the rock bridge may respond differently to different stress directions.[36] Further infinite combinations of rock-bridge configurations and stress conditions can yield a vast array of outcomes. Rock-bridge failure mechanisms can be tensile, shear, or their combination;[37] rock-bridge failure morphologies can be a single penetration, an intricate path circumventing it, or an avoidance; and rock-bridge strengths can be highly puzzling and sensitive to specific conditions.[38-41] In this context, the open question can be summarised as: which kinds of bridging structures, under what stress conditions, and to what extent they play resistance roles.

To answer this question, we have recently proposed a theory built upon the foundational concepts of Mohr's circle and failure envelope, which has been validated by experiments and applied to a rockslide and a fault linkage case.[36] Concerns about rock-bridge configuration diversity and stress-orientation dependence are decoupled, providing an operational definition of a rock bridge by deriving its resistance magnitude and failure modes, as well as the determinants, which elevates a rock bridge from a conceptual entity to an explicit input for analysing joint and fault connection processes. However, several concerns remain unresolved: (1) Incomplete abstraction into failure criteria. Previous study has remained primarily at the methodology stage, stopping short of formulating the failure criteria to serve as benchmarks for analysing rock bridges. (2) Insufficient elucidation of theoretical generalisability. While the methodology is exemplified through a rock bridge with open coplanar discontinuities subjected to unconfined compression, its generalisation to other conditions, encompassing different discontinuity arrangements (discontinuity spacing, continuity, inclination, bridging angle, etc.), different stress regimes (confined compression, tension, shear, etc.), multiple rock bridges, infilled discontinuities, and diverse field conditions (surface loading, subsurface excavation, weathering, saturation, etc.), remains insufficiently elucidated. (3) Limited experimental validation. Previous experimental validation has been limited to sandstone rock bridges in unconfined compression, lacking broader validation across different lithologies and stress regimes.

To overcome these limitations, we aim to develop rock-bridge failure criteria, which are generally applicable and can serve as benchmarks for analysing rock bridges, especially for elucidating the resistance role of rock bridges. In this paper, the rock-bridge failure criteria are introduced first, followed by their generalisation across discontinuity arrangements, stress regimes, multiple rock bridges, discontinuity infillings, and field conditions. Subsequently, we compile an extensive laboratory database of rock bridges comprising published and original datasets, and the experimental rock-bridge resistance and failure modes are compared with theoretical results. The advancement, field-scale applicability, and contribution of the rock-bridge failure criteria are highlighted, and conclusions are given finally.

## 2. Rock-bridge failure criteria

### 2.1. Failure criterion for intact rocks

Before introducing rock bridges, the failure criterion for intact rocks is briefly reviewed first. A graphical construction known as Mohr's circle can be used to visualise the relationship between normal and shear stresses ($\sigma_n$, $\tau$) acting on inclined planes at a point. Conventional unconfined and confined compressive tests on intact rocks plot principal stresses ($\sigma_1 > \sigma_2 = \sigma_3 \geq 0$) on the $\sigma_n$ axis, with circles connecting $\sigma_1$ and $\sigma_3$ to construct Mohr's circles (Fig. 1), where $\sigma_1$ and $\sigma_3$ denote the maximum and minimum principal stresses, respectively. The tangent points of these Mohr's circles form the failure envelope, which delineates the limiting shear stress a material can withstand for a given normal stress. There are multiple mathematical expressions to describe the failure envelope, among which the classical Mohr–Coulomb failure criterion is exemplified here, which is an internal frictional model suitable for brittle materials in uniaxial or low confining compression and has been widely adopted in rock mechanics. It assumes that shear failure planes at 45° + $\phi/2$ to $\sigma_3$ are developed in intact rocks. When failure occurs, the stress components on the failure planes satisfy the Coulomb shear failure criterion $\tau = \sigma_n \tan\phi + c$, where $\phi$ and $c$ denote the internal friction angle and cohesion, respectively.

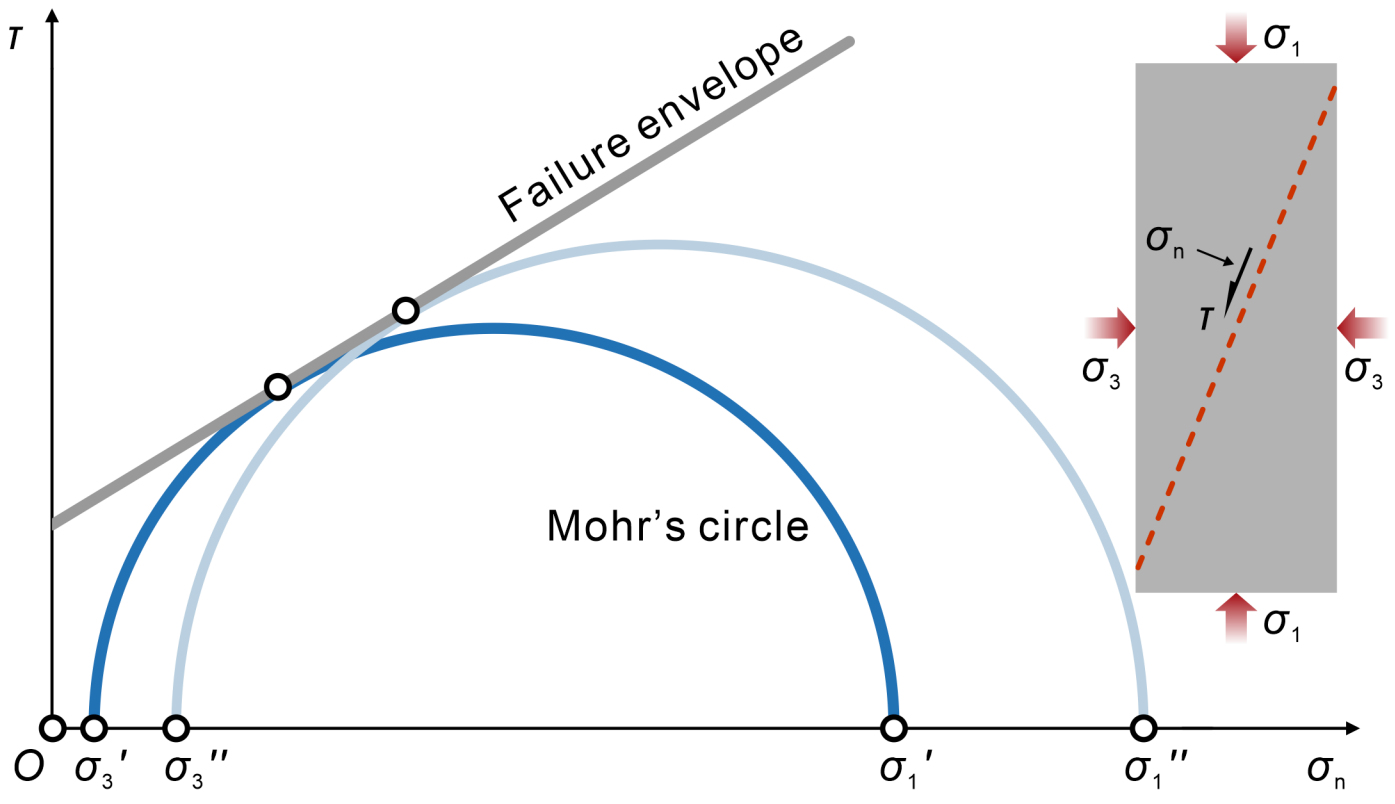

**Fig. 1.** Construction of Mohr's circles and failure envelope of intact rocks. $\tau$, shear stress; $\sigma_n$, normal stress; $\sigma_1$, maximum principal stress; $\sigma_3$, minimum principal stress.

Taking unconfined compressive tests ($\sigma_1 > \sigma_2 = \sigma_3 = 0$) on intact rocks as a further illustrative example, as shown in Fig. 2, unconfined compressive strength $\sigma_c$ and the origin construct the Mohr's circle. Given a plane aligned relative to the horizontal at $\theta$, stress components on this plane can be expressed by point P in a Mohr plane:

$$\begin{cases} \sigma_{nP} = \sigma_c \cos^2\theta \\ \tau_P = \sigma_c \sin\theta \cos\theta \end{cases} \tag{1}$$

At this moment, the Mohr's circle is tangent to the failure envelope at point R, where the stress components satisfy the Mohr–Coulomb failure criterion, that is, $\tau_R = \sigma_{nR} \tan\phi + c$. Failure will occur along the planes at 45° + $\phi/2$ to the horizontal, namely classical failure planes. In this way, both point R (stress components on classical failure planes) and point P (stress components on the given plane) correspond to the classical failure mode.

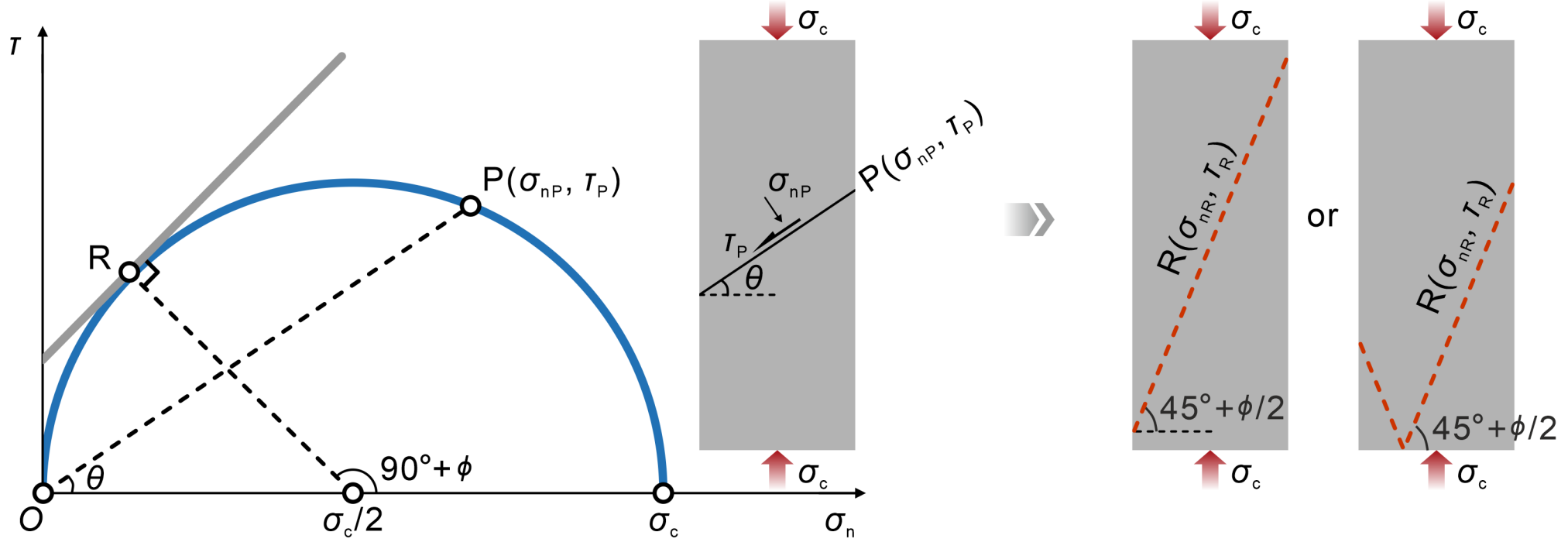

**Fig. 2.** Stress components on an inclined plane (denoted by point P) and on classical failure planes (denoted by point R) within the framework of the Mohr–Coulomb failure criterion in unconfined compression. $\sigma_c$, unconfined compressive strength; $\theta$, inclination angle of a given plane; $\phi$, internal friction angle.

### 2.2. Stress path on ligament

As exemplified by Fig. 3a, a rock bridge is defined by open coplanar discontinuities in a rectangular specimen that is clamped by loading plates and has a thickness of $b$. An external force $F$, increasing linearly with time $t$, is applied to the loading plates. In the front view (Fig. 3b), the line connecting discontinuity tips is defined as a ligament,[28] which is inclined at $\theta$ to the horizontal and has a length of $L$. Because of the dissipation of driving forces by open discontinuities, the rock bridge serves as the sole structure to transmit compression (friction between the specimen and loading plates is neglected). In other words, $F$ is directly applied to the ligament through the blue rectangle in Fig. 3b. Therefore, the stress components on the ligament can be calculated by:

$$\begin{cases} \sigma_n = \dfrac{F\cos\theta}{bL} \\ \tau = \dfrac{F\sin\theta}{bL} \end{cases} \tag{2}$$

As $F$ linearly increases, both $\sigma_n$ and $\tau$ linearly increase following $\tau = \sigma_n\tan\theta$, that is, the stress path on the ligament.

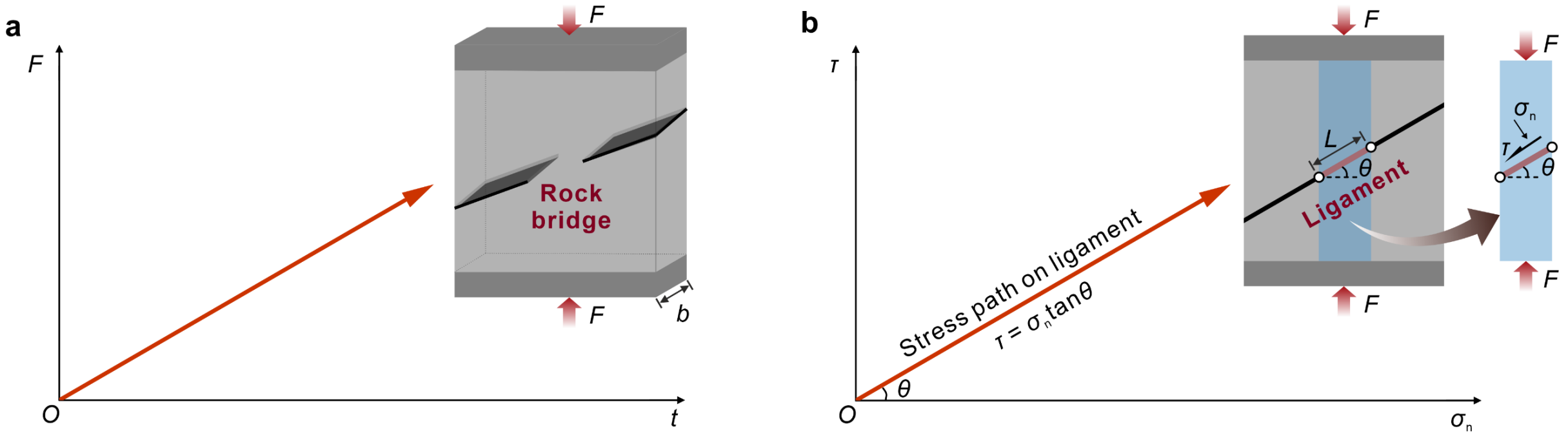

**Fig. 3.** Stress path on a ligament. (a) External force $F$ increases linearly with time $t$ and is applied to the rock-bridge specimen. (b) Calculation of the stress path on the ligament. $b$, specimen thickness; $\theta$, ligament inclination angle; $L$, ligament length.

### 2.3. Failure criteria for rock bridges defined by non-overlapping discontinuities

### 2.3.1. Classical failure

Isolating the intact rock that transmits compression, as shown in Fig. 3b, the ligament can be the given plane in Fig. 2. The stress path on the ligament can be plotted alongside Mohr's circle, forming Fig. 4a. When $\theta < 45°$ + $\phi/2$, the increasing stress path on the ligament will intersect the Mohr's circle at point P, while the Mohr's circle is tangent to the failure envelope at point R. As stated in Section 2.1, both point P and point R correspond to the classical failure mode, with classical failure planes at 45° + $\phi/2$ to the horizontal (Fig. 4b). The coordinates of point P are the simultaneous solution of the stress path on the ligament and the Mohr's circle:

$$\begin{cases} \tau = \sigma_{\mathrm{n}} \tan\theta \\ (\sigma_{\mathrm{n}} - \frac{\sigma_{\mathrm{c}}}{2})^2 + \tau^2 = (\frac{\sigma_{\mathrm{c}}}{2})^2 \end{cases} \tag{3}$$

With the derived P($\sigma_{\mathrm{nP}}$, $\tau_{\mathrm{P}}$), we can calculate the corresponding $F$ using Eq. (2):

$$F_{\mathrm{P}} = \frac{\sigma_{\mathrm{nP}} bL}{\cos\theta} \tag{4}$$

where $F_{\mathrm{P}}$ denotes the rock-bridge resistance and equals the maximum force sustained by the specimen $F_{\mathrm{max}}$.

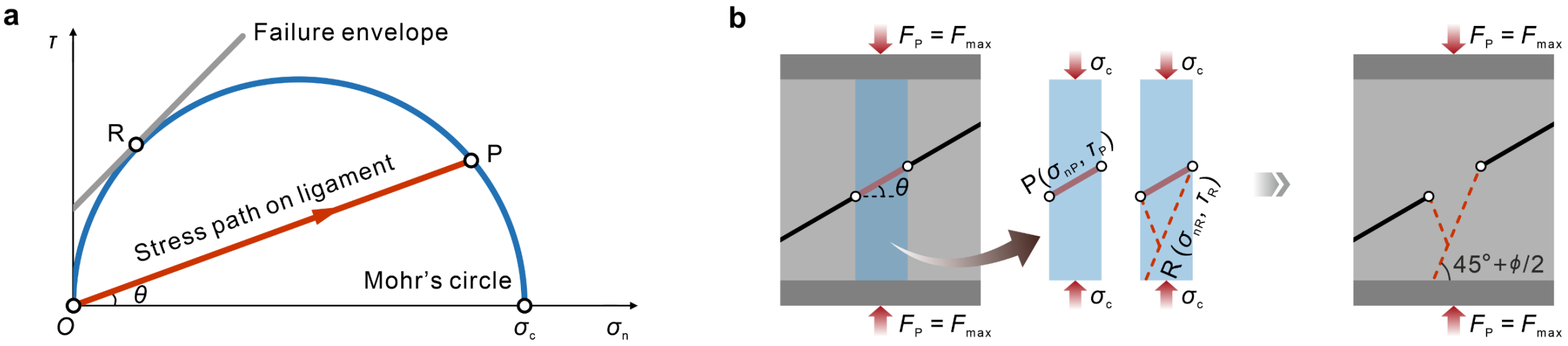

**Fig. 4.** (a) Failure criterion for classical failure. (b) Classical failure mode.

### 2.3.2. Ligament failure

When $\theta \geq 45° + \phi/2$, the increasing stress path on the ligament will also intersect the Mohr's circle at point $\mathrm{P}_1$ (Fig. 5a), at which the classical failure is expected to occur. However, the stress components on the expected classical failure planes are averaged out by the regions outside the intact rock enclosing the ligament (Fig. 5b), thus no longer reaching the limiting state, and the expected classical failure planes fail to coalesce with each other, resulting in the first intersection $\mathrm{P}_1$ no longer signifying failure. As the stress path on the ligament further increases, it will intersect the failure envelope at point $\mathrm{P}_2$ (Fig. 5a), at which the stress components on the ligament satisfy the failure criterion, indicating the failure along the ligament instead of along the classical failure plane. The coordinates of point $\mathrm{P}_2$ are the simultaneous solution of the stress path on the ligament and the failure envelope:

$$\begin{cases} \tau = \sigma_{\mathrm{n}} \tan\theta \\ \tau = f(\sigma_{\mathrm{n}}) \end{cases} \tag{5}$$

where $\tau = f(\sigma_n)$ can be exemplified by the Mohr–Coulomb failure criterion $\tau = \sigma_n \tan\phi + c$. Substituting the derived $P_2(\sigma_{nP2}, \tau_{P2})$ into Eq. (2) yields the rock-bridge resistance $F_P$:

$$F_P = \frac{\sigma_{nP2} bL}{\cos\theta} \quad (6)$$

Supplementarily, when $\theta = 90°$, $F$ is transmitted straight alongside the ligament, and no force crosses or acts upon it; thus, $F_P = 0$.

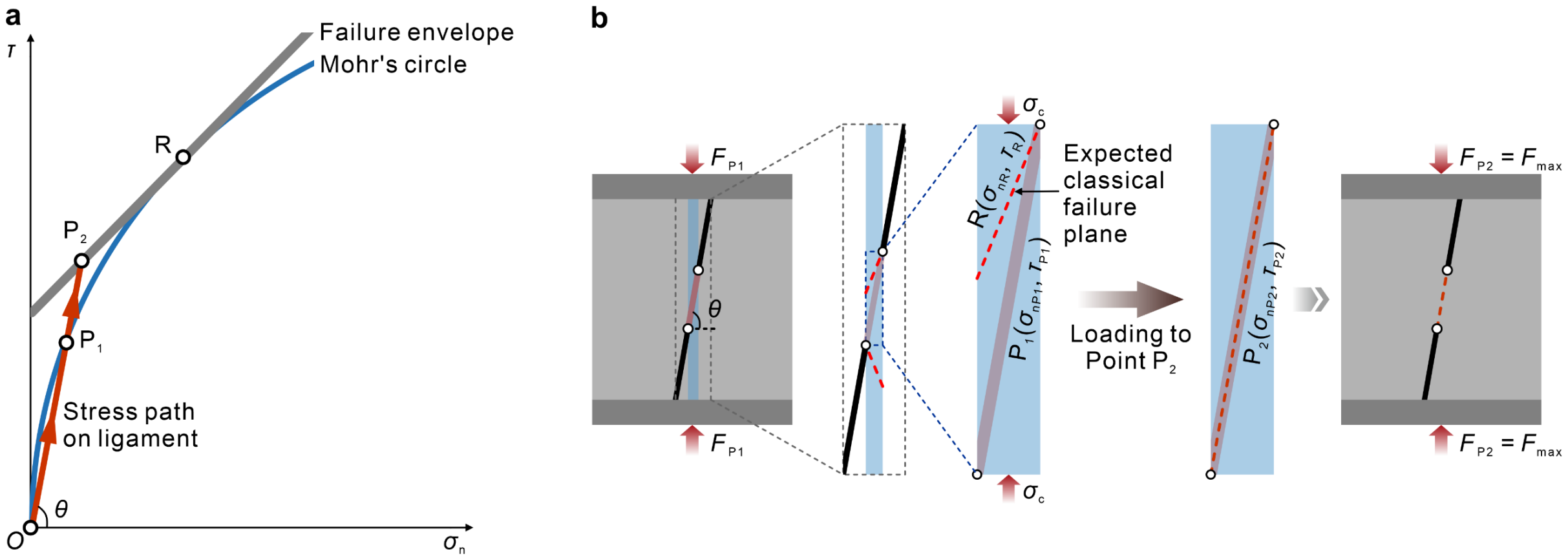

**Fig. 5.** (a) Failure criterion for ligament failure. (b) Ligament failure mode.

### 2.3.3. Failure criteria

By combining the failure criterion for classical failure and that for ligament failure, the failure criteria for rock bridges defined by non-overlapping discontinuities can be established, as shown in Fig. 6. The intersection of stress paths on ligaments with curve RA' corresponds to ligament failure, while that with curve RB corresponds to classical failure. RA' is the partial failure envelope of intact rocks, while RB is the partial Mohr's circle. They are tangent at point R, corresponding to the critical value of $\theta = 45° + \phi/2$. Both RA' and RB can be determined by basic rock mechanics tests, with Eqs. (3) and (5) calculating the coordinates of the intersection points, permitting the utilisation of curve A'RB to quantify rock-bridge resistance $F_P$.

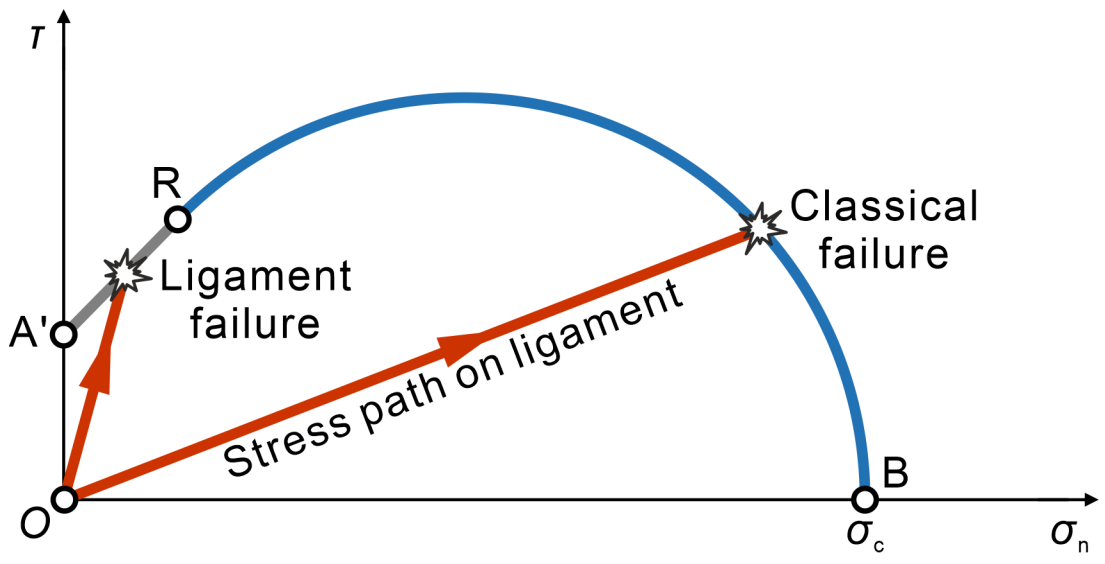

**Fig. 6.** Failure criteria for rock bridges defined by non-overlapping discontinuities subjected to unconfined compression.

## 2.4. Failure criterion for rock bridges defined by overlapping discontinuities

For a rock bridge defined by overlapping discontinuities normal to $F$, as shown in Fig. 7a, the dissipation of driving forces by open discontinuities inhibits the transmission of $F$ through the ligament and facilitates stress concentrations near discontinuity tips, generating a concentrated force $F_c$ that is approximately equal to $F$. Isolating the rectangular overlapping part with inner discontinuity tips as vertices allows it to be equivalently represented as a cantilever beam. Tension develops in the part near $F_c$, whereas compression occurs on the opposite side. The compressive-resistant but tensile-weak property of rocks prioritises the consideration of tensile stress $\sigma_{t1}$:

$$\sigma_{t1} = \frac{My}{I} \tag{7}$$

where $M$ is the bending moment induced by $F_c$, $y$ is the distance from the neutral axis, and $I$ is the moment of inertia. Substituting $M = |F_c L\cos\alpha|$, $I = bh^3/12$, and $y = h/2$ into Eq. (7), $\sigma_{t1}$ can be determined, where $h$ is the discontinuity spacing, and $\alpha$ is the bridging angle defined as $\sin^{-1}(h/L)$.

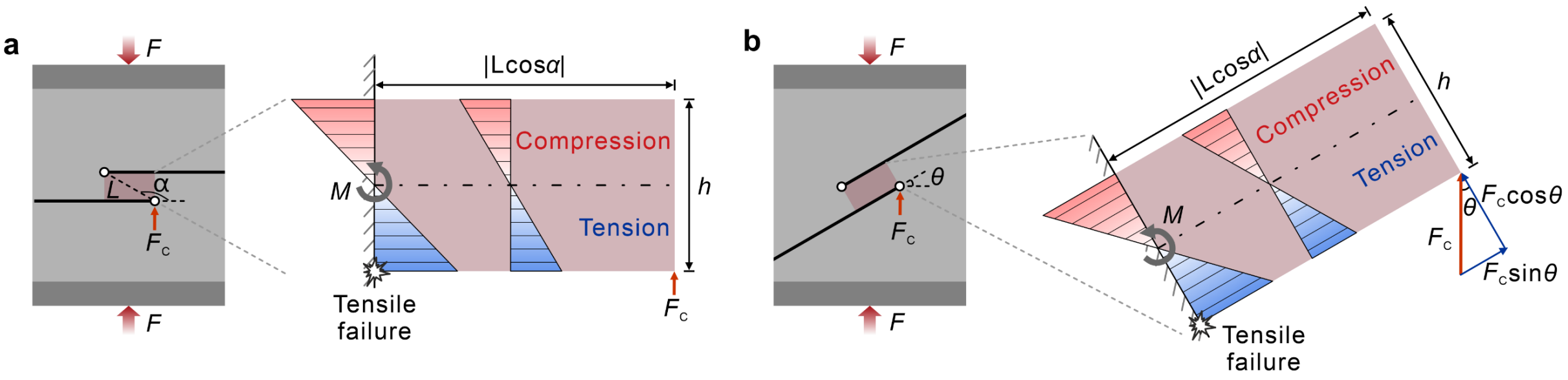

**Fig. 7.** Failure criterion for rock bridges defined by (a) horizontal; and (b) inclined overlapping discontinuities. $h$, discontinuity spacing; $\alpha$, bridging angle; $M$, bending moment; $F_c$, concentrated force.

In most cases, discontinuities and $F$ deviate from a perpendicular orientation, forming the inclination angle $\theta$, as shown in Fig. 7b. In this case, $F_c$ can be resolved into $F_c\cos\theta$ and $F_c\sin\theta$, among which $F_c\cos\theta$ results in the bending moment $M$, inducing the tensile stress $\sigma_{t1}$ that can be calculated by Eq. (7); while $F_c\sin\theta$ directly stretches the beam, inducing additional tensile stress $\sigma_{t2}$:

$$\sigma_{t2} = \frac{F_c \sin\theta}{bh} \tag{8}$$

Both $\sigma_{t1}$ and $\sigma_{t2}$ contribute to the tensile stress at the vertices beyond the inner discontinuity tips:

$$\sigma_t = \sigma_{t1} + \sigma_{t2} = \frac{6|F_c L\cos\alpha\cos\theta| + F_c h\sin\theta}{bh^2} \tag{9}$$

$\sigma_t \geq T$ serves as the failure criterion for rock bridges defined by overlapping discontinuities, where $T$ is the uniaxial tensile strength of intact rocks. Tensile failure occurs at the remaining two diagonal vertices of the rectangular overlapping part. Substituting $T$ into Eq. (9) yields the corresponding rock-bridge resistance $F_P$. Should $\sigma_t$ remain consistently lower than $T$, alternative failure modes warrant consideration.

## 2.5. Generalisation

### 2.5.1. Discontinuity arrangements

Section 2.3 takes coplanar discontinuities as an example to illustrate the rock-bridge failure criteria. For natural rock masses, the arrangements of impersistent discontinuities always deviate from coplanarity. As shown in Fig. 8a, non-overlapping discontinuities can be coplanar, oblique, and irregular to each other. In any case, geometric descriptors $L$, $\theta$, and $b$ to define rock bridges are always present. Given $L$, $\theta$, and $b$, and subjected to $F$, Eq. (2) yields the stress path on the ligament. Following the determination of curve A'RB in Fig. 6 by basic rock mechanics tests, by identifying whether the intersection of the stress path on the ligament and curve A'RB falls on curve RA' or curve RB, rock-bridge failure modes, that is, ligament failure or classical failure, can be determined. The coordinates of the intersection point that can be calculated by Eqs. (3) and (5) serve as direct inputs for estimating the resistance provided by rock bridges. In this way, the failure criteria in Section 2.3 remain compatible with oblique and irregular cases.

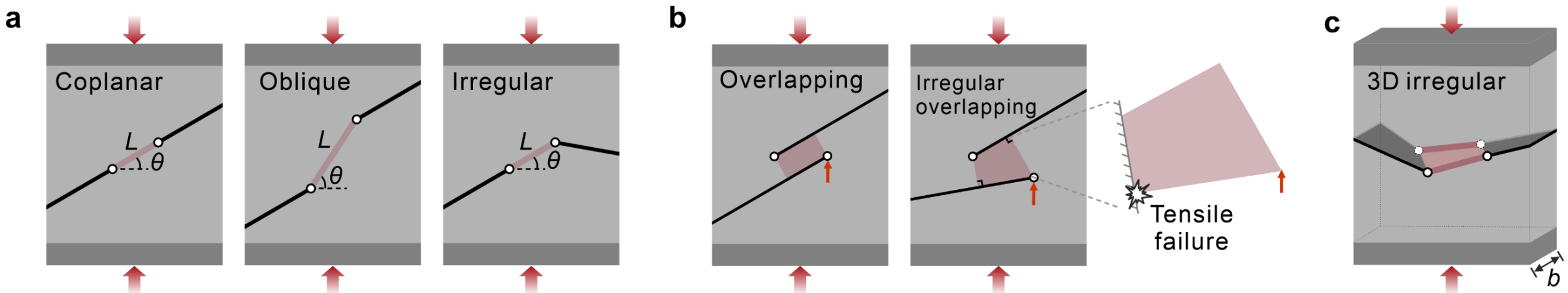

**Fig. 8.** Various discontinuity arrangements, including (a) non-overlapping; (b) overlapping; and (c) 3D irregular.

Section 2.4 takes parallel overlapping discontinuities as examples to illustrate the failure criterion for rock bridges defined by overlapping discontinuities. For non-parallel cases, as shown in Fig. 8b, the overlapping part is an irregular quadrilateral, thus requiring further calculations. Tensile failure is triggered by the sum of $\sigma_{t1}$ and $\sigma_{t2}$, in which $\sigma_{t1}$ results from the bending moment $M$ induced by $F_c\cos\theta$, and $\sigma_{t2}$ results from direct tension by $F_c\sin\theta$. The calculation of $M$ requires integrating the moment of inertia $I$ in Eq. (7) that depends on specific discontinuity arrangements. Additionally, for parallel cases, both potential tensile failure points on the diagonal fail almost simultaneously; in contrast, for non-parallel cases, the calculated tensile stresses at the two points are always different, and thus, tensile failure would occur at the point with a higher tensile stress, triggering rock-bridge failure while the other point remains intact.

These analyses are based on discontinuity traces on exposure surfaces. Naturally, discontinuity surfaces may not be parallel, as shown in Fig. 8c, while the rock-bridge failure criteria remain compatible. For non-overlapping cases, the ligament surface can always be defined by $L$, $\theta$, and $b$, based on which the stress path on the ligament surface can be determined, which can be used to analyse rock-bridge failure. For overlapping cases, tensile stresses along the two diagonal edges of the overlapping part require individual computation and comparison, according to which the failure position and rock-bridge resistance can be calculated.

### 2.5.2. Stress regimes

#### 2.5.2.1. Tension

The above analyses take unconfined compression as an example to illustrate the rock-bridge failure criteria. In tension, the failure criteria are also compatible. For rock bridges defined by non-overlapping discontinuities, subjected to tension, curve RA' is extended to RA, as shown in Fig. 9. Following the mechanisms described in Section 2.3, the intersection of stress paths on ligaments with curve RA corresponds to ligament failure, while that with curve RB corresponds to classical failure, while the $\sigma_n$ is always negative. In this case, failure will consistently occur along the ligament, and the failure criterion is expressed by Eq. (5).

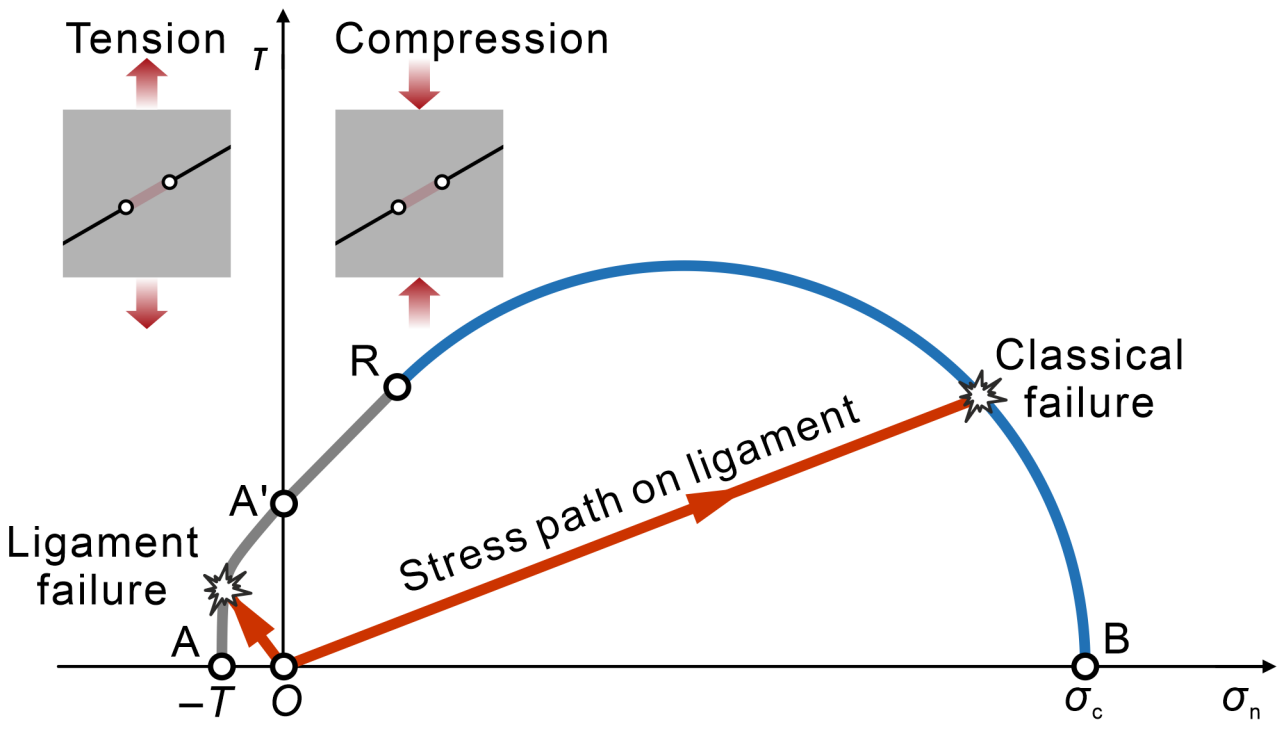

**Fig. 9.** Failure criteria for rock bridges defined by non-overlapping discontinuities in tension and compression. *T*, uniaxial tensile strength of intact rocks.

For overlapping cases, $F$ and $F_c$ in Fig. 7 turn negative in tension. $F_c$ is resolved into $F_c\cos\theta$ and $F_c\sin\theta$ but in opposite directions. Compared to the compression case, $F_c\cos\theta$ induces an equal but oppositely directed bending moment, and $F_c\sin\theta$ directly compresses the beam. Potential tensile failure position shifts from the points in Fig. 7 to the opposite diagonal vertex due to the reversal of the bending moment direction. Equation (9) is recalculated as:

$$\sigma_t = \sigma_{t1} + \sigma_{t2} = \frac{6\,|F_c L\cos\alpha\cos\theta\,|-F_c h\sin\theta}{bh^2} \tag{10}$$

$\sigma_t \geq T$ also serves as the failure criterion. $\sigma_t$ can be negative or always below $T$. In this case, alternative failure modes can be possible.

#### 2.5.2.2. Triaxial compression

The failure criteria for rock bridges defined by non-overlapping discontinuities are also compatible with biaxial and triaxial compression, as well as more complex boundary conditions. Conventional triaxial compression ($\sigma_1 > \sigma_2 = \sigma_3 > 0$) is selected as an example to illustrate. Compared to Fig. 3, the minor compression force $F_3$ is applied to a predetermined value and is kept constant, after which the major compression force $F_1$ is applied and linearly increases (Fig. 10a).

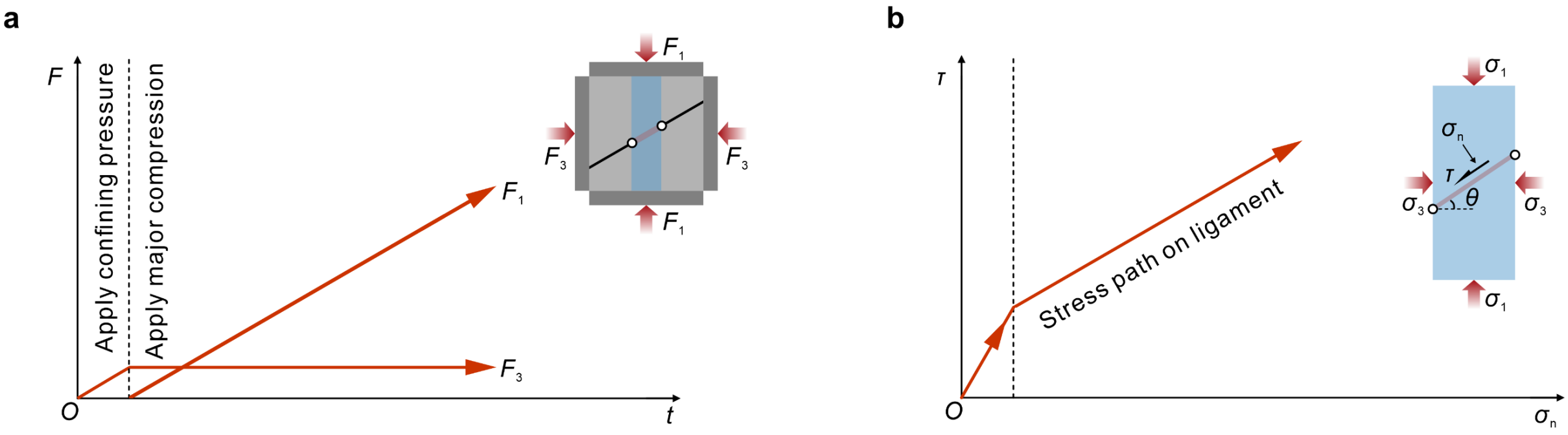

**Fig. 10.** Stress path on a ligament in conventional triaxial compression. (a) Force path. (b) Stress path.

As shown in Fig. 10b, $F_3$ results in an initial stress state on the ligament. With a linearly increasing $F_1$ and constants $\theta$, $F_3$, $b$, and $L$, the stress path on the ligament increases linearly, which can be expressed by:

$$\tau = \sigma_n \frac{F_1 \sin\theta + F_3 \cos\theta}{F_1 \cos\theta + F_3 \sin\theta} \tag{11}$$

As the stress path on the ligament increases, the corresponding Mohr's circle enlarges and touches the failure envelope at point R, and the stress path intersects the Mohr's circle at point P (Fig. 11a). According to the mechanism described in Section 2.3.1, classical failure will occur, forming classical failure planes at 45° + $\phi$/2 to $F_3$ (Fig. 11b). The coordinates of point P are the simultaneous solution of the stress path on the ligament and the Mohr's circle:

$$\begin{cases} \tau = \sigma_n \dfrac{F_1 \sin\theta + F_3 \cos\theta}{F_1 \cos\theta + F_3 \sin\theta} \\ (\sigma_n - \dfrac{\sigma_1 + \sigma_3}{2})^2 + \tau^2 = (\dfrac{\sigma_1 - \sigma_3}{2})^2 \end{cases} \tag{12}$$

With the derived P($\sigma_{nP}$, $\tau_P$), rock bridge resistance $F_P$ can be calculated.

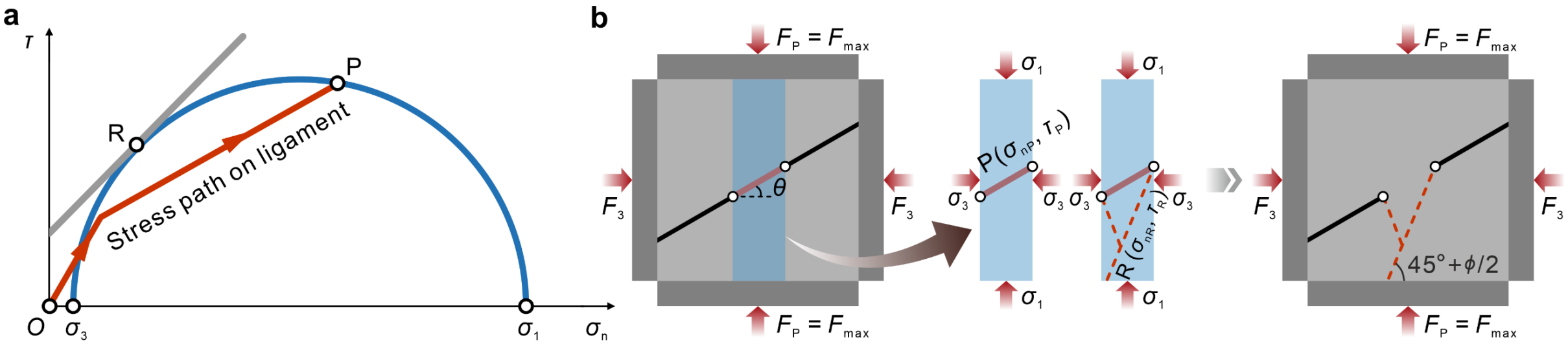

**Fig. 11.** (a) Failure criterion for classical failure in conventional triaxial compression. (b) Classical failure mode.

With a higher slope of the stress path on the ligament, as shown in Fig. 12a, the stress path intersects the Mohr's circle at point $P_1$, while classical failure will not occur. As the stress path further intersects the failure envelope at point P$_2$, ligament failure will occur (Fig. 12b). See Section 2.3.2 for the mechanisms. The coordinates of point P$_2$ are the simultaneous solution of the stress path on the ligament and the failure envelope:

$$\begin{cases} \tau = \sigma_{\mathrm{n}} \dfrac{F_1 \sin\theta + F_3 \cos\theta}{F_1 \cos\theta + F_3 \sin\theta} \\ \tau = f(\sigma_{\mathrm{n}}) \end{cases} \tag{13}$$

With the derived P($\sigma_{\mathrm{nP}}$, $\tau_{\mathrm{P}}$), rock-bridge resistance $F_{\mathrm{P}}$ can be calculated.

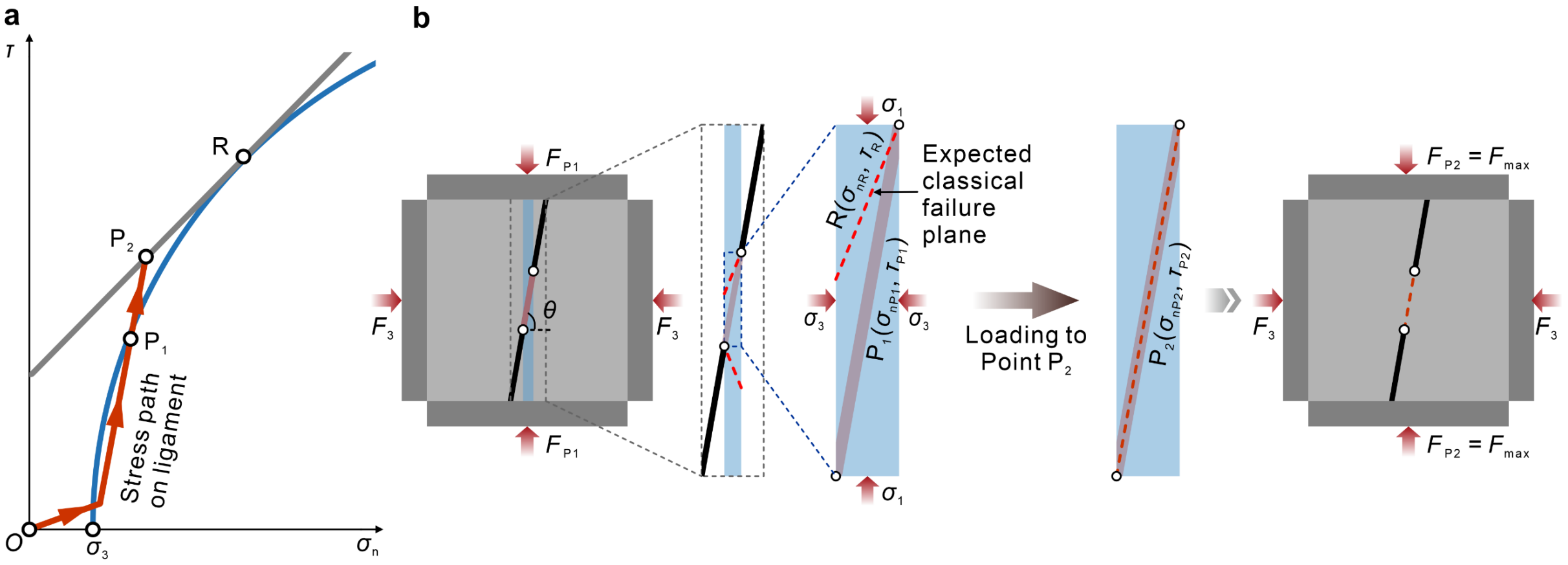

**Fig. 12.** (a) Failure criterion for ligament failure in conventional triaxial compression. (b) Ligament failure mode.

Failure criteria for rock bridges defined by non-overlapping discontinuities across stress regimes can be summarised in Fig. 13. Compared to Fig. 9, both the position of Mohr's circle and stress paths on ligaments change, while the concept to identify whether stress paths on ligaments intersect curve RA or curve RB remains compatible.

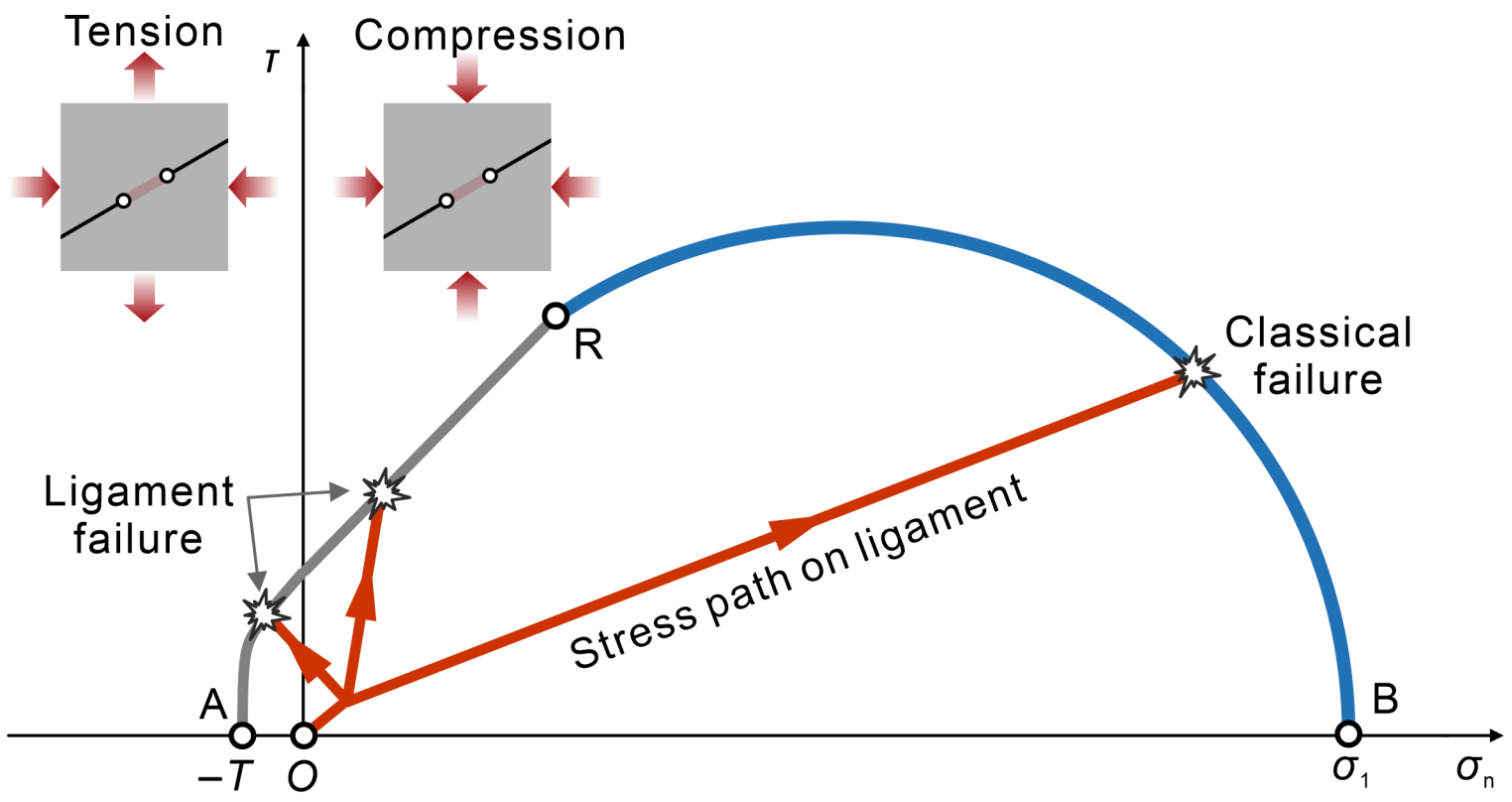

**Fig. 13.** Failure criteria for rock bridges defined by non-overlapping discontinuities across stress regimes.

For overlapping cases, the cantilever beam simplification is also applicable under complex stress regimes, while the calculation of $\sigma_{\mathrm{t}}$ can refer to Eqs. (9) and (10).

### 2.5.3. Multiple rock bridges

Natural rock masses may contain multiple rock bridges. As exemplified by Fig. 14, rock-bridge mapping yields ligament information. After boundary conditions are applied, the stress path on each ligament can be obtained, where the arithmetic-averaging approach employed throughout the preceding analysis is not recommended, as the external load may not simply be distributed; instead, the digital image correlation (DIC) conversion, numerical simulation, and analytical solution, as detailed in Section 4.2, provide reliable alternatives to obtain stress paths on ligaments. Subsequently, the rock-bridge failure criteria can be applied bridge by bridge. The one whose ligament stress path first intersects curve ARB is determined to fail first, and the intersection point on curve RA or RB determines its resistance and failure modes including ligament failure or classical failure. Then, the analysis proceeds to the next rock bridge. As the failure of the first rock bridge may induce stress redistribution, stress paths on ligaments should be updated, after which the rock-bridge failure criteria are applied again to predict the failure behaviour. This process is repeated until the rock mass fails, where the rock-bridge failure criteria yield the failure sequence, resistance magnitude, and failure modes of these rock bridges. Note that the collective resistance of the rock mass should be treated through identifying the critical bridge (or bridges) from individual rock-bridge resistance $F_P$; because of stress redistribution, the total resistance is not a simple sum of individual $F_P$. When the rock bridges defined by non-overlapping and overlapping discontinuities coexist, the above procedures by applying the rock-bridge failure criteria bridge by bridge remain applicable.

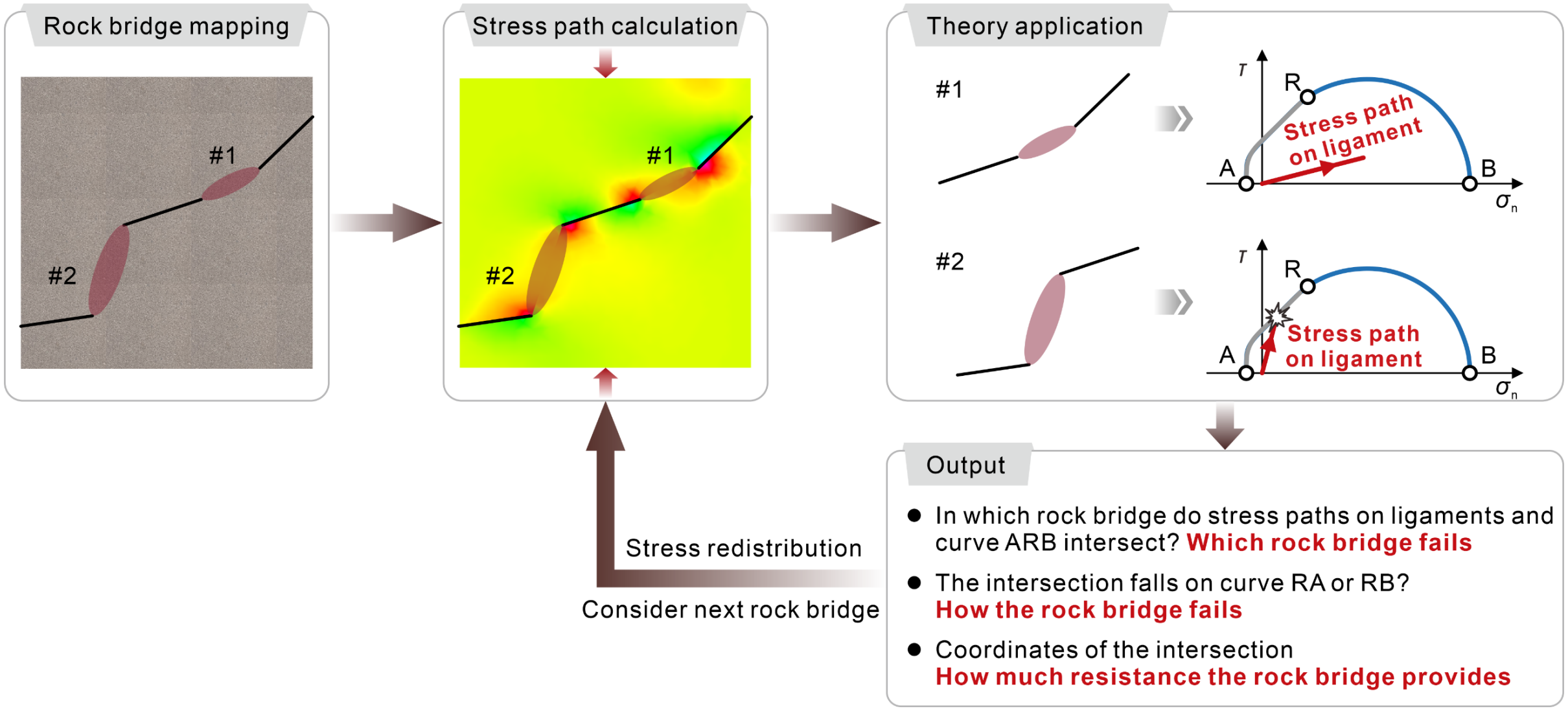

**Fig. 14.** Applicability to multiple rock bridges.[36]

#### 2.5.4. Discontinuity infillings

The above analyses focus on rock bridges with open discontinuities, where rock bridges serve as the sole structures for load transmission. In some cases, discontinuities may be infilled by sand, clay, silt, breccia, gouge, or rock fragments, thereby sharing the load with rock bridges. Failure of discontinuities can be described by the classical Coulomb friction criterion:

$$\tau_{\mathrm{in}} = c_{\mathrm{j}} + \sigma_{\mathrm{n_in}} \tan\phi_{\mathrm{j}} \tag{14}$$

where $\tau_{in}$ and $\sigma_{n_in}$ represent the shear and normal stresses acting on discontinuities, $c_j$ and $\phi_j$ denote the cohesion and friction angle of discontinuities. For non-overlapping cases, $\tau_{in}$ reaching the shear strength of infilling materials before the intersection of stress paths on ligaments and curve ARB implies preferential discontinuity slippage, leaving theoretically unaffected intersection coordinates. In this case, the rock-bridge failure criteria remain compatible, and rock-bridge resistance and failure modes can be derived from the coordinates of the intersections. Conversely, if $\tau_{in}$ remains below the shear strength of infilling materials at the intersection of stress paths on ligaments and curve ARB, rock-bridge failure would be suppressed by static friction from discontinuities. With further increasing $F$ until $\tau_{in}$ reaches the limiting state, discontinuities would slide, accompanied by the rock-bridge failure.

For overlapping cases, both the $M$-induced tension $\sigma_{t1}$ and the direct tension $\sigma_{t2}$ can be affected by the infill. For $\sigma_{t1}$, when the stiffness of the infill is much lower than that of the rock matrix, Eq. (7) still holds; however, when the infill is stiffer than the rock matrix, such as iron-siliceous materials, $M$ is suppressed, as well as the $\sigma_{t1}$. For $\sigma_{t2}$, the friction from discontinuities inhibits the direct tension; therefore, $\tau_{in}$ should be subtracted in Eq. (8).

#### 2.5.5. Field conditions

For brevity, the above analyses assume linearly increasing external loads and fixed Mohr's circles or failure envelopes in the failure criteria for rock bridges defined by non-overlapping discontinuities. Natural rock mass instability is a multifaceted process. Here we consider four representative potential factors, including surface loading, excavation, weathering, and water intrusion.

Additional surface construction implies surface loading. As shown in Fig. 15a, the initial Mohr's circle (dashed blue curve) remains far away from the failure envelope, indicating stability. The stress components on the ligament inclined at $\theta$ to the horizontal are represented by point P. With increasing surface loading, the Mohr's circle will enlarge and touch the failure envelope at point R, and the stress components on the ligament will evolve to point P'. According to the rock-bridge failure criteria, we can determine whether failure will occur along the ligament or along classical failure planes, with the coordinate of point P' to further calculate the maximum force sustained by the rock bridge.

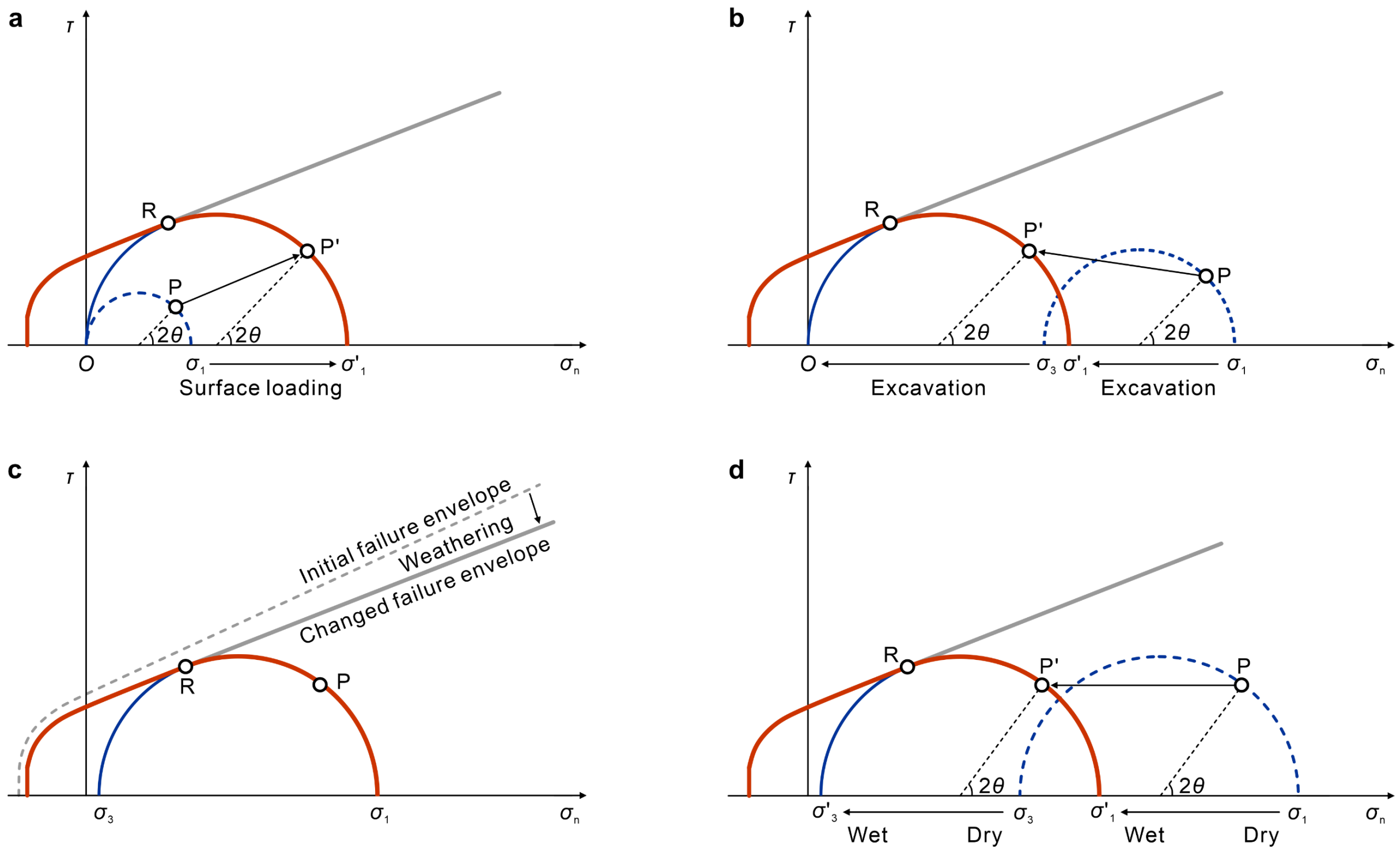

**Fig. 15.** Guidelines for generalising the failure criteria for rock bridges defined by non-overlapping discontinuities in the field, encompassing (a) surface loading; (b) excavation; (c) weathering; and (d) water intrusion.

Surface or subsurface excavation results in stress redistribution. As shown in Fig. 15b, the *in situ* rock mass is initially in a triaxial stress state. After excavation, due to the man-made free surface, the local stress state may change to uniaxial or biaxial stress states. The *in situ* stress components on the ligament, denoted by point P, will evolve to point P', and the corresponding Mohr's circle will change its position. If the Mohr's circle is tangent to the failure envelope, instability will occur. Failure modes and morphologies of rock bridges can be deduced by analysing point P with the rock-bridge failure criteria.

Weathering on rocks weakens their mechanical properties, such as $c$ and $\phi$, thereby lowering the failure envelope, as shown in Fig. 15c. As weathering progresses, the changed failure envelope will touch the Mohr's circle that represents the *in situ* stress state of the rock mass. By incorporating the stress components on the ligament, represented by point P, into the rock-bridge failure criteria, failure modes and morphologies of rock bridges can be determined.

The intrusion of water into rocks increases pore pressure, thereby reducing the effective stress. As shown in Fig. 15d, $\sigma_1$ and $\sigma_3$ will decrease, shifting the Mohr's circle to the left. During this process, the stress components on the ligament denoted by point P will shift simultaneously to those denoted by point P'. When the shifted Mohr's circle touches the failure envelope, the rock-bridge failure criteria can be used to analyse rock-bridge failure.

For other factors contributing to rock mass instability, the principle of generalising the rock-bridge failure criteria is similar. These factors are often coupled, necessitating an integrated analysis. On the other hand, the failure criterion for overlapping cases would be easier to apply, requiring consideration of whether tensile stress increases or the tensile strength of rocks decreases in given field conditions.

## 3. Experimental validation

### 3.1. Experimental database

#### 3.1.1. Data from the literature

Extensive research on rock bridges has preset discontinuities with varying quantities and configurations in a simulated infinite medium subjected to specific stresses, as shown in Fig. 16a. Discontinuity arrangements define rock-bridge configurations, which can be described by ligament length $L$, discontinuity spacing $h$, bridging angle $\alpha$, and discontinuity inclination angle $\beta$, where $\alpha = \sin^{-1}(h/L)$, and $\alpha + \beta = \theta$ (Fig. 16b). However, in this setup, intact surrounding rocks on both sides (red prisms in Fig. 16c) also serve as rock bridges but are typically neglected, yet their higher stiffness compared to the middle flawed part (white prism in Fig. 16c) presents a potential impact on the results. In other words, the independent assessment of the interested rock bridge would be hindered because of surrounding rocks (rock bridges). To solve this problem, prefabricated edge discontinuities are introduced (Fig. 16d), where a single rock bridge serves as the sole structure to transmit load. Nevertheless, the edge discontinuities may intersect the loading surface (Fig. 16e), which also brings the effect of surrounding rocks. To exclude the abovementioned interference, this study restricts the experimental database to Fig. 16d.

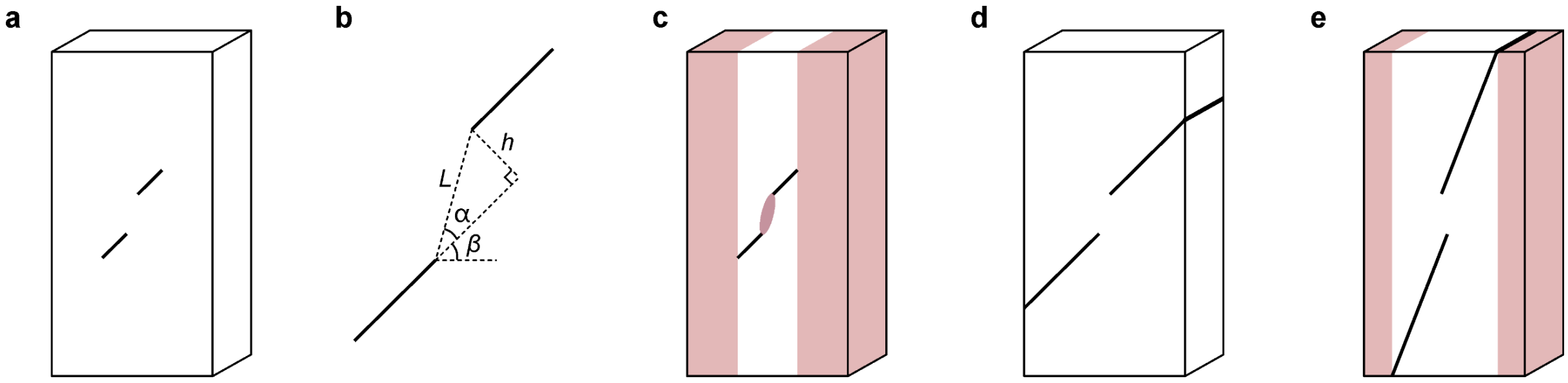

**Fig. 16.** Lab-scale rock bridge settings. Taking a prism rock specimen as an example, the vast majority of studies on rock bridges adopt (a) intact rock embedded with several prefabricated discontinuities with (b) geometric descriptors ($L$, ligament length; $h$, discontinuity spacing; $\alpha$, bridging angle; $\beta$, discontinuity inclination angle) to define rock bridges, which (c) neglects the role of surrounding rocks (red prisms), hindering an independent assessment of the rock bridge. (d) Prefabricated edge discontinuities provide an optimal solution for independently assessing rock bridges, yet (e) surrounding rocks remain influential when edge discontinuities and loading surfaces intersect.

We manually reviewed the literature on rock bridges from 1962 to 2025, based on which we compiled an experimental database on testing rock bridges using the settings in Fig. 16d. The experimental database comprises fifty-two unconfined compressive tests and ninety-four confined compressive tests, with discontinuities in forty of them being filled by mica or ice. Rock (sandstone, granite, and marble) and rock-like materials (cement mortar) were cut into cylindrical or prismatic shapes to be quasi-statically loaded in a lab. Discontinuities are arranged in various patterns, not merely in parallel, reflecting a wide range of possible scenarios.

### 3.1.2. Data from our experiments

Following the literature review, we identified several issues: (1) Insufficient experimental data for overlapping cases. To date, experimental designs with overlapping edge discontinuities remain limited. (2) Absence of consideration of stress-orientation dependence. The established experimental database contains experimental designs involving various geometric descriptors of rock bridges $L$, $h$, $\alpha$, and $\beta$. Nevertheless, with identical geometric descriptors, rock bridges may perform differently when subjected to different stress directions.

To address these issues, we prefabricated coplanar, oblique, and overlapping open discontinuities in prismatic sandstones, which were subjected to unconfined compression. As the rotation of stress directions is equivalent to that of rock bridges, stress directions and geometric descriptors $L$, $h$ and $\alpha$ were fixed, while rock bridges were anticlockwise rotated by angles of $\beta$ = 0°, 30°, 45°, 60°, and 90° from the horizontal (Fig. 17). Centroids of rock bridges were the same as the front view centres of specimens. In this way, identical rock bridges were exposed to stresses in different directions. See Bu et al.[36] for more experimental details.

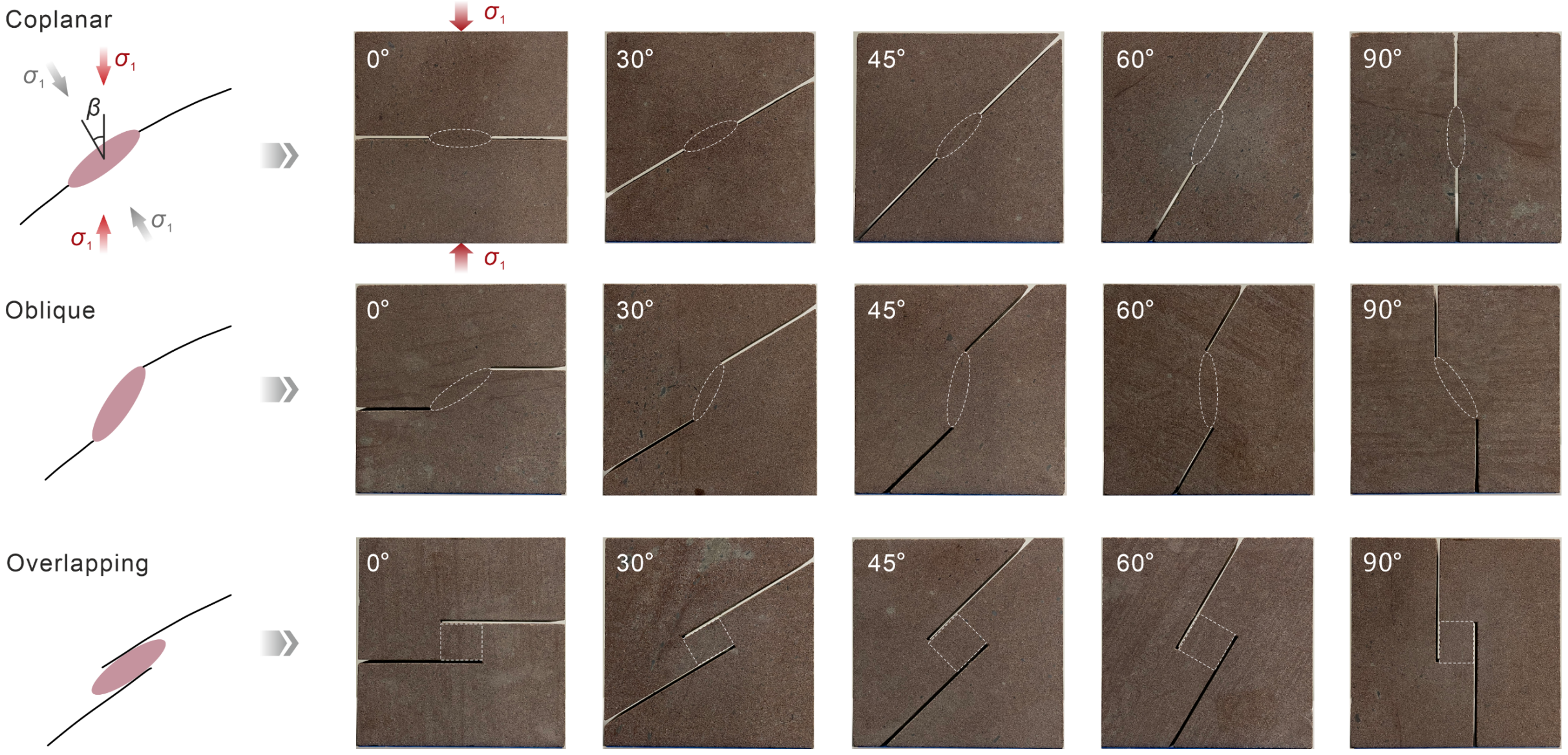

**Fig. 17.** Prefabricated lab-scale rock bridges under controlled geometries and stress orientations. For each category, geometric descriptors of rock bridges remain constant, while $\beta$ serves as the sole variable.

### 3.2. Comparison between experimental and theoretical results

### 3.2.1. Rock-bridge resistance

Figure 18 compares experimental and theoretical rock-bridge resistance $F_P$. Specimens are subjected to unconfined compression. The comparison results can be summarised as follows:

(1) With all mean relative errors remaining below 0.15, the feasibility of the rock-bridge failure criteria to deduce rock-bridge resistance $F_P$ is demonstrated.

(2) The rock-bridge failure criteria are not restricted to specific lithologies, as demonstrated across the sandstone,[42-44] marble,[44, 45] and granite[44, 46] examined here. The unconfined compressive strength $\sigma_c$ here ranges from 31.29 to 186.89 MPa, covering the $\sigma_c$ range of most lithologies.

(3) The rock-bridge failure criteria are applicable to various specimen shapes. Most are cylinders with a height-to-diameter ratio of two, while prisms with height-to-length ratios of two[47] and one[44] also exhibit strong applicability.

(4) The rock-bridge failure criteria are compatible with rock bridges defined by non-parallel discontinuities. Hao et al.[44] set different edge-discontinuity arrangements, and the deviation between theoretical and experimental $F_P$ is within 0.04 for non-parallel discontinuities, which validates the generalisation in Section 2.5.1.

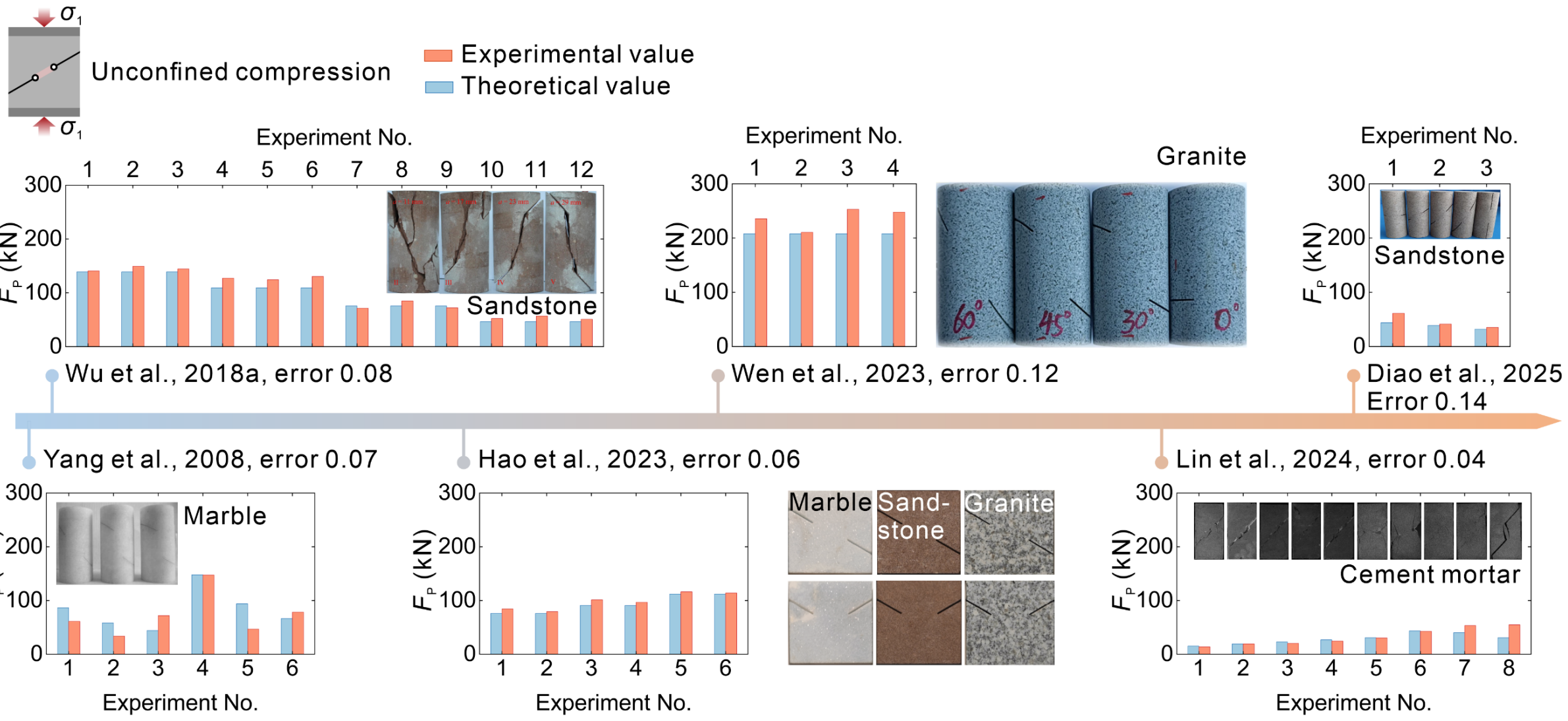

**Fig. 18.** Comparison of experimental and theoretical rock-bridge resistance $F_P$ in unconfined compression.

To further validate the rock-bridge failure criteria in confined compression, $F_P$ acquired by previous experiments and the rock-bridge failure criteria is compared in Fig. 19. The results are summarised below:

(1) The mean relative error between experimental and theoretical $F_P$ is always below 0.23, demonstrating the validity of the rock-bridge failure criteria under confined compression.

(2) The rock-bridge failure criteria are applicable across various stress conditions. The testing schemes include conventional triaxial[15, 41, 48-50] and biaxial loading forms,[51] with the former being predominant, while the latter also yield an agreement between experimental and theoretical $F_P$, where the mean relative error is 0.19. The experimental database contains sandstones and granites with $\sigma_c$ of 77.34–296.82 MPa, friction angles of 28–56.9°, and cohesion of 18.52–70.39 MPa, subjected to confining pressures of 0–40 MPa. Under different combinations of lithology and confining pressure, the rock-bridge failure criteria consistently predict the experimental results.

The comparison indicates that the generalisation of the failure criteria in Section 2.5.2 is valid.

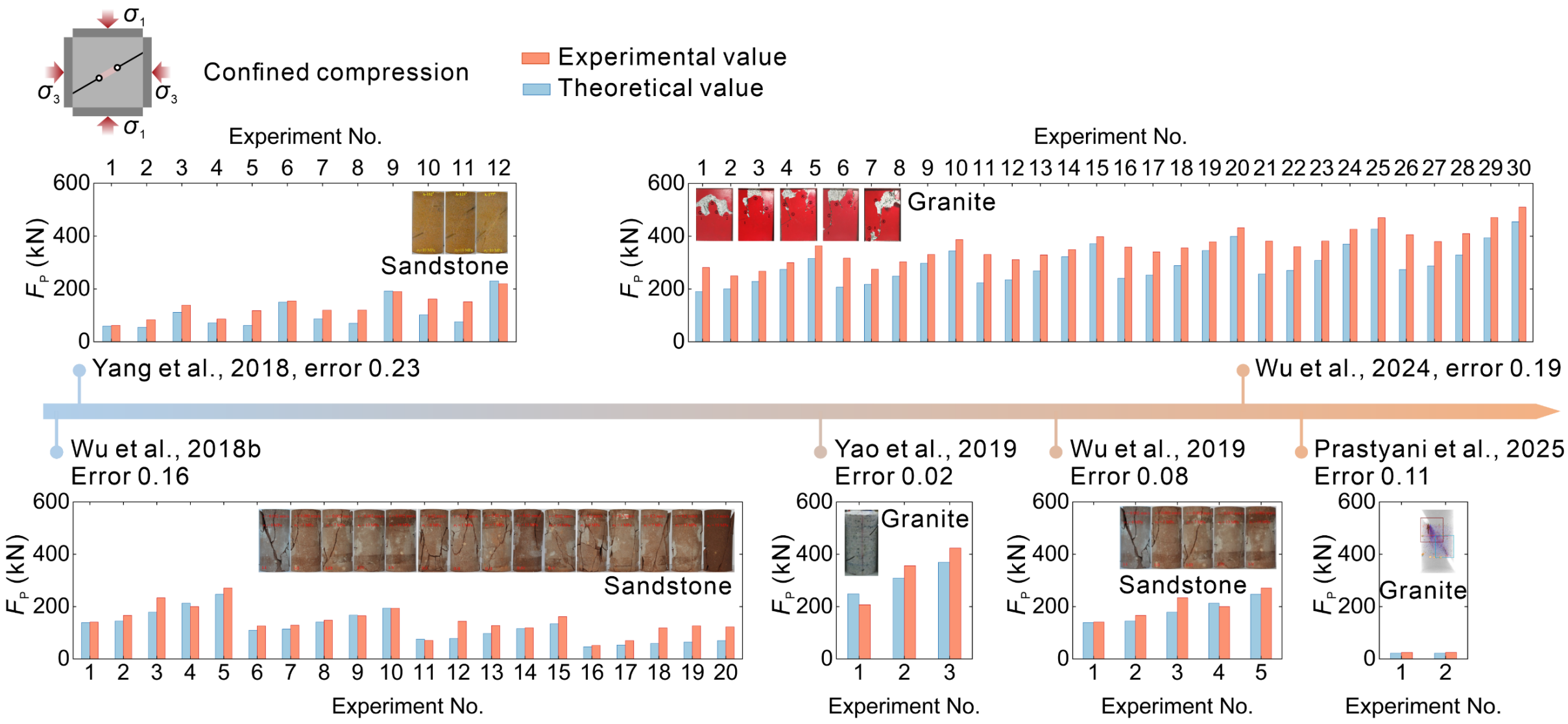

**Fig. 19.** Comparison of experimental and theoretical $F_P$ in confined compression.

Some researchers prefabricated rock bridges with infilled edge discontinuities, in contrast to the above-discussed rock bridges with open discontinuities. Figure 20 compares the experimental and theoretical $F_P$, from which the following results are obtained:

(1) The mean relative error between the experimental and theoretical $F_P$ is within 0.18, demonstrating the validity of the generalisation of the failure criteria in Section 2.5.4.

(2) The failure criteria are applicable to cylindrical sandstones with ice-filled discontinuities subjected to confined compression[52] and prismatic ones with mica-filled discontinuities under unconfined compression.[53] Both ice and mica exhibit low strength and stiffness, thereby exerting little effect on rock-bridge roles.

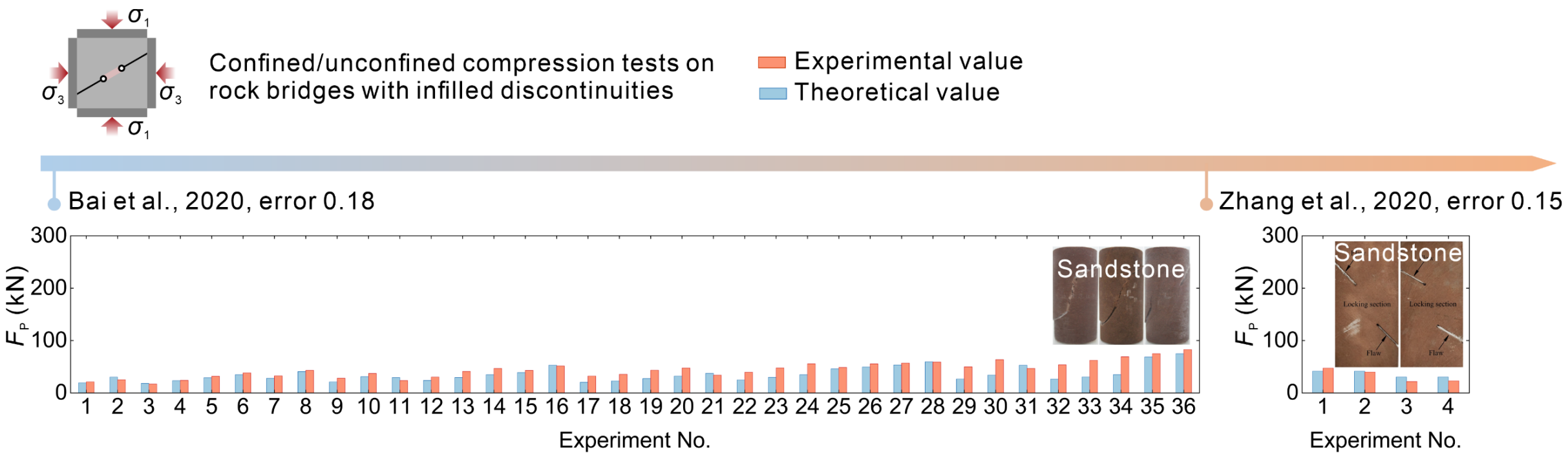

**Fig. 20.** Comparison of experimental and theoretical $F_P$ with infilled discontinuities subjected to compression.

Figure 21 compares experimental $F_P$ from our experiments with theoretical $F_P$ from the rock-bridge failure criteria. Consistency between experimental and theoretical $F_P$ is supported by the coefficient of determination $R^2$ of 0.94. Here, we selected $R^2$ as the metric rather than the mean relative error, as $R^2$ is dimensionless and scale-invariant, allowing it to effectively evaluate the overall goodness of fit for the datasets of coplanar, oblique, and overlapping cases without being biased by the difference in their orders of magnitude. Note that certain $F_P$ values are directly measured by stress sensors, as they equal the maximum force sustained by the specimen, whereas others are reconstructed according to digital image correlation (DIC) results (cases with boxed $\beta$ in Fig. 21) because the specimen dimensions necessitate the inclusion of surrounding rocks like Fig. 16e to ensure the equivalence between rotating stress directions and rock bridges. For the directly measured $F_P$, the $R^2$ value against theoretical predictions reaches 0.97. Moreover, Fig. 21c indicates that the rock bridges defined by overlapping discontinuities present a considerably narrower $F_P$ range (0–3 kN) compared to the non-overlapping cases (0–94 kN). Cases with $\beta$ = 60° and 90° are not displayed because the resistance role of rock bridges is significantly suppressed by surrounding rocks, and the failure deviated from that suggested by the cantilever beam simplification.

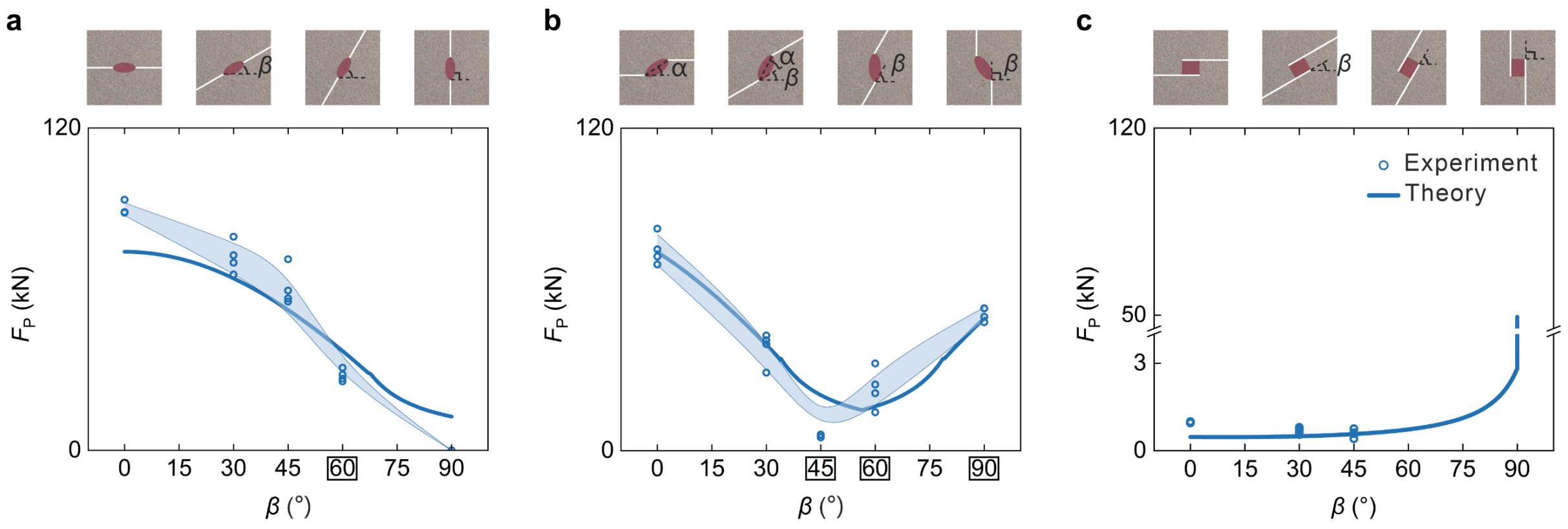

**Fig. 21.** Comparison of experimental and theoretical $F_P$ of (a) coplanar; (b) oblique; and (c) overlapping cases in our experiments.[36] Shaded areas connected by splines denote the standard deviations of experimental $F_P$. Boxed

$\beta$ values correspond to the case using the DIC to reconstruct $F_P$, whereas unboxed $\beta$ values denote those directly measuring $F_P$.

### 3.2.2. Failure morphology

In addition to the rock-bridge resistance $F_P$, the rock-bridge failure criteria also provide an analytical solution for rock-bridge failure morphology, including classical failure, ligament failure, and cantilever beam failure. Based on the partial experimental database, the failure morphology observed in experiments and predicted by the failure criteria is compared in Fig. 22. Since the failure morphology in the experimental database is displayed as 2D images, it can be affected by the camera angle, especially for cylindrical specimens owing to their curved surfaces. Consequently, instead of comparing the precise location of the failure morphology such as the rupture angle, this comparison focuses on the categories into which the failure morphology falls, namely classical failure, ligament failure, and cantilever beam failure.

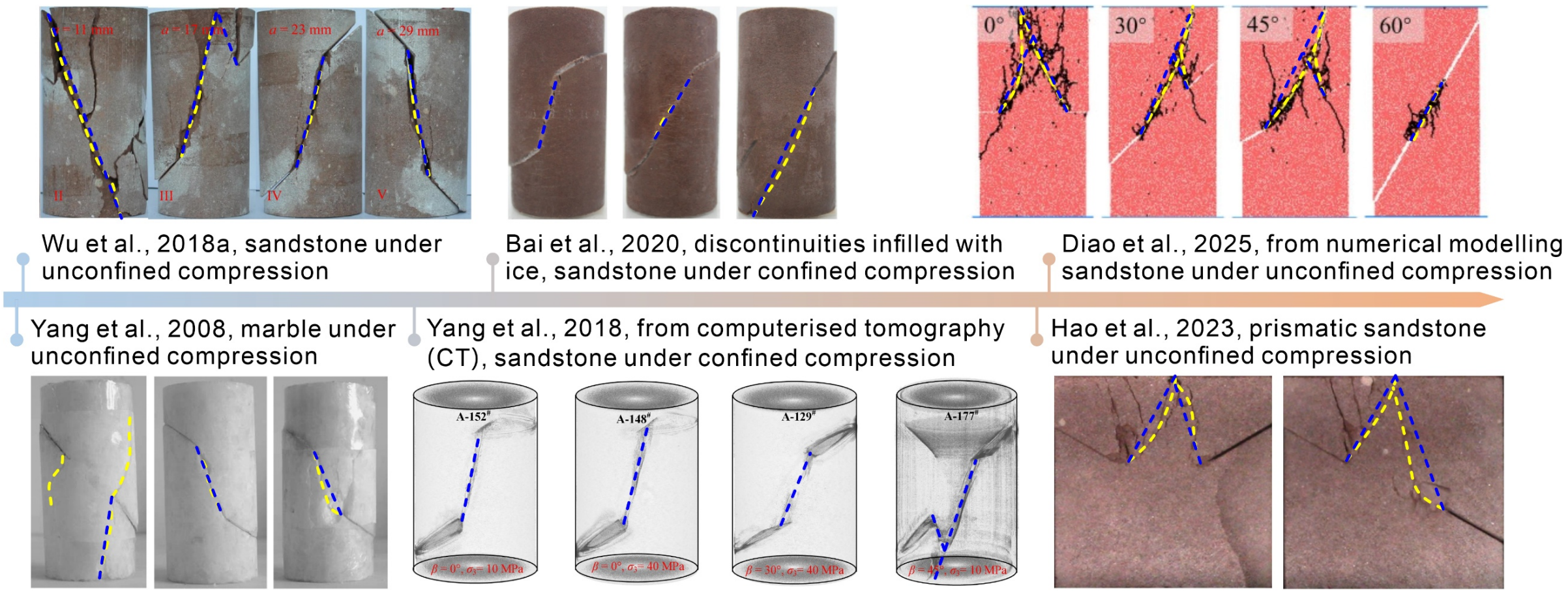

**Fig. 22.** Comparison of experimental and theoretical failure morphologies, where yellow and blue dashed lines denote experimental and theoretical failure morphologies, respectively.

As shown in Fig. 22, experimental failure morphologies generally agree with those predicted by the rock-bridge failure criteria. Failure observations of cylindrical sandstone and marble specimens show that some catastrophic ruptures follow a nearly straight line along ligaments, while others deviate from ligaments and exhibit classical failure morphology. Both classical failure and ligament failure are correctly predicted by the rock-bridge failure criteria. Numerical simulations may provide an ideal failure morphology of rock bridges. The discrete element modelling of sandstone bridges under unconfined compression yields the failure morphology that closely agrees with the theoretical prediction.[43]

For overlapping cases, the cantilever beam simplification suggests that instability manifests as tensile failure at the vertices beyond the inner discontinuity tips within the rectangular overlapping part. As shown in Fig. 23, the observed failure morphology in our experiment all conforms to the theoretical prediction. Two tensile cracks initiated at the diagonal vertices far from the inner discontinuity tips; then, they propagated towards these tips, forming the anticlockwise rotation of the overlapping part.

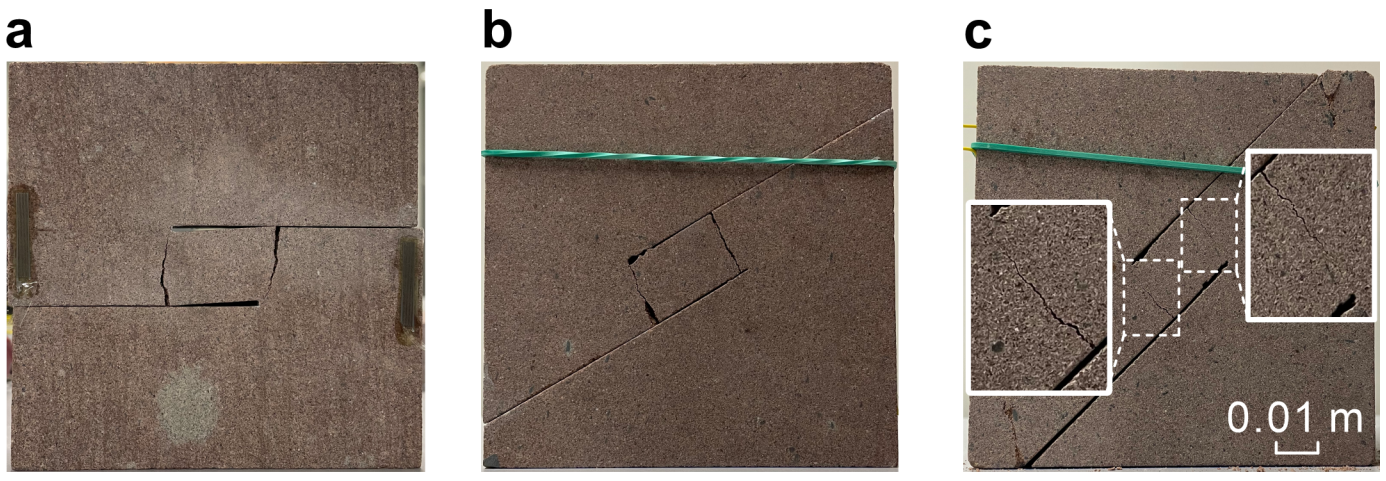

**Fig. 23.** Experimental failure morphology of the rock-bridge specimen containing overlapping discontinuities. (a) $\beta$ = 0°; (b) $\beta$ = 30°; and (c) $\beta$ = 45°.

## 4. Discussion

### 4.1. Advancement

We also evaluated our experimental data against the prevailing Jennings's criterion[21] and its coupling with Jaeger's single plane of weakness theory.[25] In contrast to the robust experimental agreement achieved by the rock-bridge failure criteria ($R^2$ = 0.94), they yield inadequate predictions with $R^2$ of 0.21 and 0.02, respectively. Such a deficiency stems mainly from two fundamental limitations: (1) Oversimplified geometric description of rock bridges. Conventional models rely heavily on discontinuity persistence *K*, which fails to describe oblique and overlapping discontinuities, along with other limitations such as the omission of progressive failure and scale effects,[29, 30] thus necessitating reliable geometric descriptors. (2) Limited failure mechanism. Since rock-bridge failure can be tensile, shear, or their combination,[37] analysing rock-bridge failure through a preconceived singular mode is fundamentally inadequate.

The rock-bridge failure criteria address these two limitations. (1) We replace *K* with geometric descriptors focusing on the ligament, including ligament length *L* and ligament inclination angle $\theta$, which are not limited to specific rock-bridge configurations. (2) We determine rock-bridge failure mechanisms by identifying the intersection of stress paths on ligaments and Mohr's circle or failure envelope, along with the cantilever beam simplification, covering the possible tensile, shear, and combined failure mechanisms. In addition, the failure criteria have their origins in classical rock mechanics theories. Specifically, the failure criteria for non-overlapping cases stem from the foundational concepts of Mohr's circle and failure envelope, that is, Mohr's circle visualises the stress components on ligaments, and both the Mohr's circle touching the failure envelope and the failure envelope itself indicate failure.

Another limitation in conventional models is the assumption of simultaneous failure of all rock bridges in rock masses. Our model instead shows that failure initiates sequentially at rock bridges where the local stress path on the ligament intersects curve ARB, defining rock mass failure as a progressive process.

### 4.2. Field-scale applicability

To analyse field-scale rock bridges using the rock-bridge failure criteria, rock-bridge mapping is required to obtain the essential inputs. Previous mapping has focused on estimating the rock-bridge percentage, namely 1 – *K*, as summarised by Tuckey and Stead[54] in landslide research. However, as discussed in Section 4.1, *K* fails to constrain rock-bridge configurations. Therefore, what should be mapped is the discontinuity arrangement, from which ligament information, including $L$, $\theta$, and $b$, is derived, rather than 1 – *K*.

Despite identifying the target parameters, their precise measurement can be difficult. Current technology is difficult to precisely map the subsurface discontinuities within large-scale natural rock masses. Most researchers still rely on limited outcrop information, where discontinuity length is assumed to be the diameter of a disc or the edge of a polygon inscribing the mapped feature. Regardless of the assumed geometry, the mapped discontinuity remains an explicit 2D measure of the trace intersecting a geological exposure. Consequently, this linear trace does not necessarily correspond directly to the underlying discontinuity, as evidenced by the natural discontinuity and rock-bridge profiles by Shang et al.[55] Alternative methods include borehole drilling, ground-penetrating radar, ultrasonic testing, infrared thermal imaging, etc. For example, we employed infrared thermal imaging to measure rock-bridge configurations in the potential Boot Flake rockslide in an exfoliating cliff of El Capitan within Yosemite Valley, California, USA. See Guerin et al.[34] and Bu et al.[36] for more details.

The inputs should also include the curve ARB in Fig. 9, where RA is the partial failure envelope of intact rocks, and RB is the partial Mohr's circle. Following field sampling, conventional unconfined and confined compressive tests, along with Brazilian splitting tests or direct tensile tests, should be conducted to determine the failure envelope and the Mohr's circle. Such a low parameter demand promotes the practical application of the rock-bridge failure criteria.

Following the abovementioned inputs, the determination of stress paths on ligaments is the next step. To illustrate our concept straightforwardly, this study assumes linearly increasing external loads, and the stress paths on ligaments are determined using the idealised arithmetic average approach, as shown in Eq. (2). This can be suitable for analysing the lab-scale single rock bridge. However, in applications of the rock-bridge failure criteria, such as field-scale rock bridges and rock masses containing multiple rock bridges, it may fail to capture the stress paths on ligaments. Because stress distribution can be non-uniform and be affected by multiple factors, including the evolution of stress conditions due to the overburden and tectonic stresses, the deterioration or reinforcement process, spatio-temporal dynamics of hydrothermal conditions, fault or rock-bridge interactions, weathering, and human activities. This does not indicate the inapplicability of the rock-bridge failure criteria, but rather necessitates robust methods to obtain the stress paths on ligaments. Alternative methods include: (1) Analytical method. Küntz et al.[56, 57] proposed analytical methods for calculating the stress fields in cracked media by establishing the correspondence between equilibrium equations for an incompressible elastic solid and a viscous fluid in steady state, from which the stress paths on ligaments can be obtained. (2) Experimental method. DIC enables the extraction of strain fields that can be converted into stress fields using Hooke's law, which further yields stress paths on ligaments. Bu et al.[36] used this method to obtain the experimental stress paths on ligaments, which is also compatible with multiple rock bridges. (3) Numerical method. Numerical modelling of rock masses allows the stress fields to be extracted;[58, 59] subsequently, the stress paths on ligaments can be obtained. This method was used in both the potential Boot Flake rockslide and the fault linkage between the Southern San Andreas Fault and Imperial Fault in Bu et al.[36]

Given the stress paths on ligaments and curve ARB, the rock-bridge failure criteria can be applied. Beyond outputting the relationships between stress states and strength like most failure criteria, we further derive the rock-bridge resistance $F_P$ using Eq. (4) and focus our experimental validation on $F_P$. Because rock bridges are intact rocks, and their strengths can be equivalent to the intrinsic material strength under certain conditions. By comparison, $F_P$ takes into account $L$, $b$, and $\theta$, thereby additionally incorporating rock-bridge configurations and stress orientations, and can therefore be used to elucidate the resistance role of rock bridges. In applications, $F_P$ may receive more attention. For instance, when identifying the critical rock bridge among multiple rock bridges, their strengths can be very similar, while their $F_P$ values vary and can directly reveal which rock bridge is critical.

Regarding application scenarios, Section 2.5.5 generalises the rock-bridge failure criteria across four field conditions, including surface loading, excavation, weathering, and water intrusion, while our previous study demonstrates their practical significance in analysing landslides and fault hard-linkage. Evidently, the rock-bridge failure criteria are not restricted to isolated scenarios but offer general applicability for analysing rock mass instabilities.

### 4.3. Contribution of the rock-bridge failure criteria

Since the conceptualisation of rock bridges by Müller[3] and Terzaghi,[4] despite over half a century of extensive research on this topic, mainstream geoscientific thought treats rock bridges as conceptual entities rather than analysable objects. For example, Elmo et al.[29] proposed the first important and yet unanswered question in the rock-bridge domain: 'What is a rock bridge and what parameters govern whether a given intact rock portion of the rock mass can or should be defined as a rock bridge', as well as 'How can and should we incorporate rock bridges in stability calculations'.

To address this issue, the rock-bridge failure criteria start by clarifying that rock bridges are purely geological structures. From a geometric perspective, defining a rock bridge as a portion of intact rock between tips or edges of adjacent impersistent discontinuities is considered acceptable. However, what we are concerned with is its mechanical impact on stability, calling for moving beyond the geometry-based definition to establish a corresponding mechanical interpretation. Regretfully, the relationship between rock-bridge configurations and mechanical behaviour remains elusive. Therefore, even if rock-bridge configurations are precisely mapped, their mechanical impact on the stability remains unknown. This does not imply that rock bridges do not exist or are unmeasurable, but rather that their mechanical roles are yet to be assessed.

The rock-bridge failure criteria clarify the relationship between rock-bridge configurations and mechanical behaviour, as detailed in Sections 2 and 4.1. In this way, the rock-bridge failure criteria serve as benchmarks for analysing rock bridges, elevating rock bridges from conceptual entities to analysable objects, thereby advancing our understanding and management of the associated geological phenomena and rock engineering.

## 5. Conclusions

Rock bridges have long been recognised as pivotal in resisting rock mass instabilities. However, which kinds of rock bridges, under what stress conditions, and to what extent they play resistant roles, remain open questions, as the relationship between rock-bridge configurations and mechanical behaviour remains elusive. To address this issue, we develop rock-bridge failure criteria, yielding analytical solutions for rock-bridge resistance and failure modes, which are further compared with experimental results. The following key conclusions can be drawn:

(1) The rock-bridge failure criteria are theoretically sound. The failure criterion for rock bridges defined by non-overlapping discontinuities stems from the foundational concepts of Mohr's circle and failure envelope, while that for overlapping cases builds upon the cantilever beam simplification.

(2) The rock-bridge failure criteria are generally applicable. It has been demonstrated that they can be generalised and applied across various discontinuity arrangements, stress regimes, multiple rock bridges, infilled discontinuities, and field conditions, as the core concept remains compatible.

(3) The rock-bridge failure criteria are experimentally validated. Both the rock-bridge resistance and failure modes acquired from the rock-bridge failure criteria and experiments present a high level of agreement.

(4) The rock-bridge failure criteria hold considerable promise in practice. Following field mapping on rock bridges, the failure criteria can provide answers regarding the extent to which the rock bridge contributes to resistance, the mechanism by which it fails, and the failure morphology.

The rock-bridge failure criteria provide a fundamental solution for analysing rock bridges by establishing the relationship between rock-bridge configurations and mechanical behaviour, elevating rock bridges from conceptual entities to analysable objects.

**Acknowledgements**

This study was financially supported by the National Natural Science Foundation of China grants 41977249 and 42090052, the Swiss National Science Foundation Postdoc.Mobility fellowship P500-2_241983, and the China Scholarship Council grant 202204910040. Authors extend their appreciation to H.H. Einstein at MIT, B.D. Collins at USGS, V. Merrien-Soukatchoff at CNAM, and M. Grenon at ULaval for the discussions and comments.

**Author contributions**

F.B. conceived the idea, proposed the theory, analysed the theoretical, experimental, numerical, and field data, and wrote the manuscript. M.J. and L.X. discussed the results and finalised the manuscript. All the authors contributed to the comments on the manuscript.

**Competing interests declaration**

The authors declare no competing interests.

**Data availability**

Data supporting the findings of this study are available from the corresponding author upon a reasonable request.

**Correspondence and requests for materials** should be addressed to Fengchang Bu.